\documentstyle[12pt]{article}

\begin{document}

\input amssym.tex

\title{Dirac fermions in de Sitter and anti-de Sitter backgrounds}

\author{Ion I. Cot\u aescu\\ {\it West University of Timi\c soara,}\\{\it V.
Parvan Ave. 4, RO-300223 Timi\c soara}}

\maketitle

\begin{abstract}
Starting with a new theory of symmetries generated by isometries in
field theories with spin, one finds the generators of the spinor
representation in backgrounds with a given symmetry. In this manner
one obtains a collection of conserved operators from which one can
chose the complete sets of commuting operators defining quantum
modes. In this framework, the quantum modes of the free Dirac field
on de Sitter or anti-de Sitter spacetimes can be completely derived
in static or moving charts. One presents the discrete quantum modes,
in the central static charts of the anti-de Sitter spacetime, whose
eigenspinors can be normalized. The consequence is that the second
quantization can be done in this case in canonical manner. For the
free Dirac field on de Sitter manifolds this can not be done in
static charts being forced to consider the moving ones. The quantum
modes of the free Dirac field in these charts are used for writing
down the quantum Dirac field and its one-particle operators.

Pacs: 04.20.Cv, 04.62.+v, 11.30.-j
\end{abstract}
\

\newpage

\section{Introduction}

In general relativity \cite{SW,MTW,WALD} the development of the
quantum field theory in curved spacetimes \cite{BD} give rise to
many difficult problems related to the physical interpretation of
the one-particle quantum modes that may indicate how to quantize
the fields. This is because the form and the properties of the
particular solutions of the free field equations \cite{SOL1} -
\cite{LO} are strongly dependent on the procedure of separation of
variables and, implicitly, on the choice of the local chart
(natural frame). Moreover, when the fields  have spin the
situation is more complicated since then the field equations and,
therefore, the form of their particular solutions depend, in
addition, on the tetrad gauge in which one works \cite{UK,SW}. In
these conditions it would be helpful to use the traditional method
of the quantum theory in flat spacetime based on the complete sets
of commuting operators that determine the quantum modes as common
eigenstates and give physical meaning to the constants of the
separation of variables which are just the eigenvalues of these
operators.

A good step in this direction could be to proceed like in special
relativity looking for the generators of the geometric symmetries
similar to the familiar momentum, angular momentum and spin
operators of the Poincar\' e covariant field theories \cite{W}.
However, the relativistic covariance  in the sense of general
relativity is too general to play the same role as the Lorentz or
Poincar\' e covariance in special relativity. In its turn the
tetrad gauge invariance of the theories with spin represents
another kind of general symmetry that is not able to produce
itself conserved quantities \cite{SW}. Therefore, one must focus
only on the isometry transformations that point out the specific
spacetime symmetry related to the presence of Killing vectors
\cite{SW,WALD,ON}.

Another important  problem is how to define the generators of
these representations for any spin when some spin parts could
appear. In the case of the Dirac field these spin parts were
calculated not only in some particular cases \cite{DGB} but even
in the general case of any generator corresponding to a Killing
vector in any chart and arbitrary tetrad gauge fixing \cite{CART}.
Starting with this important result, we have generalized this
theory for any spin, formulating the theory of {\em external
symmetry} for any curved manifold \cite{C2}.

Our approach is a general theory of tetrad gauge invariant fields
defined on curved spacetimes with given external symmetries. This
predicts how must transform these fields under isometries in order
to leave invariant the form of the field equations and to obtain
the general form of the generators of these transformations. The
basic idea is that the isometries transformations must preserve
the position of the local frames with respect to the natural one.
Such transformations can be constructed as isometries {\em
combined} with suitable tetrad gauge transformations necessary for
keeping unchanged the tetrad field components. In this way  we
obtain the external symmetry group showing that it is locally
isomorphic with the isometry group.

Moreover, we define the operator-valued representations of the
external symmetry group carried by spaces of fields with spin. We
point out that these are induced  by the linear finite-dimensional
representations of the $SL(2,\Bbb C)$ group. This is the motive
why the symmetry transformations which leave invariant the field
equations have generators with a composite structure. These have
the usual orbital terms of the scalar representation and, in
addition, specific spin terms which depend on the choice of the
tetrad gauge even in the case of the fields with integer spin.  In
general, the spin and orbital terms do not commute to each other
apart from some special gauge fixings where the fields transform
manifestly covariant under external symmetry transformations.

As examples we study the central symmetry and the maximal symmetry
of the de Sitter (dS) and anti-de Sitter (AdS) spacetimes. In the
central charts we define a suitable version of Cartesian tetrad
gauge which allowed us recently to find new analytical solutions of
the Dirac equation \cite{COTA, COTA1}. We show that in this gauge
fixing the central symmetry becomes global and, consequently, the
spin parts of its generators are the same as those of special
relativity \cite{BJD,TH}. For the dS and AdS spacetimes we calculate
the generators of any representation of the external symmetry group
in central charts with our Cartesian gauge.

Furthermore, we show how can be used these results for finding the
quantum modes of the Dirac field on AdS and dS spacetimes. First we
consider (static) central charts where the Dirac equation can be
analytically solved. On AdS spacetime we obtain only quantum
discrete modes of given energy described by fundamental solutions
that can be correctly normalized. Unfortunately, in the case of the
dS spacetime we find fundamental solutions corresponding to a
continuous energy spectrum that can not be normalized in the
generalized sense. In these circumstances we are able to quantize
the Dirac field in central charts only in the AdS background.
However, normalized solutions of the Dirac equation on dS spacetime
can be found in {\em moving} charts with Cartesian \cite{C3} or
spherical coordinates \cite{C4}. For  quantizing  the Dirac field it
is convenient to chose the moving charts with Cartesian coordinates
where we can identify the components of the momentum operator and
normalize the fundamental solutions using the momentum
representation \cite{C3}. Obviously, to this end our theory of
external symmetry is crucial.

We start presenting in the second section the basic ideas of the
relativistic and gauge covariance which will be embedded in our
theory of external symmetry in the next section. The fourth section
is devoted to the charts with central symmetry while the dS and AdS
symmetries are discussed in section 5. The main features of the
Dirac theory in curved manifolds are reviewed in the next section.
Section 7 is devoted to the quantum modes of the Dirac field in
central charts of AdS and dS spacetime. We show that the second
quantization of this field can be performed in canonical manner only
in AdS case. For quantizing the Dirac field in dS spacetimes  we
must choose moving charts where we have a suitable momentum
representation. The theory of the Dirac field in these charts and
the quantization procedure are presented in section 8. Useful
formulas are given in Appendices.

We work in natural units with $\hbar=c=1$.

\section{Relativistic covariance}

In the Lagrangian field theory in curved spacetimes the relativistic
covariant equations of scalar, vector or tensor fields arise from
actions that are invariant under general coordinate transformations.
Moreover, when the fields have spin in the sense of the $SL(2,\Bbb
C)$ symmetry then the action must be invariant under tetrad gauge
transformations \cite{UK}. The first step to our approach  is to
embed both these types of transformations into new ones, called
combined transformations, that will help us to understand the
relativistic covariance in its most general terms.

\subsection{Gauge transformations}

Let us consider the curved spacetime $M$ and a local chart
(natural frame) of coordinates $x^{\mu}, \mu=0,1,2,3$. Given a
gauge, we denote by $e_{\hat\mu}(x)$ the tetrad fields that define
the local (unholonomic) frames, in each point $x$, and by $\hat
e^{\hat\mu}(x)$ those defining the corresponding coframes. These
fields have the usual orthonormalization properties
\begin{equation}\label{(duale)}
\hat e^{\hat\mu}_{\alpha}\, e_{\hat\nu}^{\alpha}=\delta^{\hat\mu}_{\hat\nu}
\,,\quad
\hat e^{\hat\mu}_{\alpha}\, e_{\hat\mu}^{\beta}=\delta^{\beta}_{\alpha}
\,,\quad
e_{\hat\mu}\cdot e_{\hat\nu}=\eta_{\hat\mu \hat\nu}\,, \quad
\hat e^{\hat\mu}\cdot \hat e^{\hat\nu}=\eta^{\hat\mu \hat\nu}\,,
\end{equation}
where $\eta=$diag$(1,-1,-1,-1)$ is the Minkowski metric. From the
line element
\begin{equation}\label{(met)}
ds^{2}=\eta_{\hat\mu \hat\nu}d\hat x^{\hat\mu}d\hat x^{\hat\nu}=
g_{\mu \nu}(x)dx^{\mu}dx^{\nu}\,,
\end{equation}
expressed in terms of 1-forms,
$d\hat x^{\hat\mu}=\hat e_{\nu}^{\hat\mu}dx^{\nu}$, we get the
components of the metric tensor of the natural frame,
\begin{equation}\label{gmunu}
g_{\mu \nu}=\eta_{\hat\alpha\hat\beta}\hat e^{\hat\alpha}_{\mu}\hat
e^{\hat\beta}_{\nu}\,,\quad
g^{\mu \nu}=\eta^{\hat\alpha\hat\beta} e_{\hat\alpha}^{\mu}
e_{\hat\beta}^{\nu}\,.
\end{equation}
These raise or lower the {\em natural} vector indices, i.e., the
Greek ones ranging from 0 to 3, while for the {\em local} vector
indices, denoted by hat Greeks and having the same range, we must
use the Minkowski metric. The local derivatives
$\hat\partial_{\hat\nu}=e^{\mu}_{\hat\nu}\partial_{\mu}$ satisfy
the commutation rules
\begin{equation}
[\hat\partial_{\hat\mu},\hat\partial_{\hat\nu}]
=e_{\hat\mu}^{\alpha} e_{\hat\nu}^{\beta}(\hat e^{\hat\sigma}_{\alpha,\beta}-
\hat e^{\hat\sigma}_{\beta,\alpha})\hat\partial_{\hat\sigma}
=C_{\hat\mu \hat\nu
\cdot}^{~\cdot \cdot \hat\sigma}\hat\partial_{\hat\sigma}
\end{equation}
defining the Cartan coefficients which help us to write the {\em
conecttion} components in local frames as
\begin{equation}
\hat\Gamma^{\hat\sigma}_{\hat\mu \hat\nu}=e_{\hat\mu}^{\alpha}
e_{\hat\nu}^{\beta}(\hat e_{\gamma}^{\hat\sigma}
\Gamma^{\gamma}_{\alpha \beta}
-\hat e^{\hat\sigma}_{\beta, \alpha})=
\frac{1}{2}\eta^{\hat\sigma \hat\lambda}(C_{\hat\mu \hat\nu \hat\lambda}+
C_{\hat\lambda \hat\mu \hat\nu}+C_{\hat\lambda \hat\nu \hat\mu})\,.
\end{equation}
We specify that this connection is often called {\em spin}
connection (and denoted by $\Omega^{\hat\sigma}_{\hat\mu
\hat\nu}$) in order to do not be confused with the Christoffel
symbols, $\Gamma^{\gamma}_{\alpha \beta}$, involved in the
well-known formulas of the usual covariant derivatives
$\nabla_{\mu}=~_{;\mu}$\,.

The Minkowski metric $\eta_{\hat\mu\hat\nu}$ remains invariant under
the transformations of the {\em gauge} group of this metric,
$G(\eta)=O(3,1)$. This has as subgroup the Lorentz group,
$L_{+}^{\uparrow}$, of the transformations $\Lambda[A(\omega)]$
corresponding to the transformations $A(\omega)\in SL(2,\Bbb C)$
through the canonical homomorphism \cite{W}. In the standard {\em
covariant} parametrization, with the real parameters
$\omega^{\hat\alpha \hat\beta}=-\omega^{\hat\beta\hat\alpha}$, we
have
\begin{equation}\label{Aomega}
A(\omega)= e^{-\frac{i}{2}\omega^{\hat\alpha\hat\beta}
S_{\hat\alpha\hat\beta}}\,,
\end{equation}
where $S_{\hat\alpha \hat\beta}$ are the covariant basis-generators of the
$SL(2,\Bbb C)$ Lie algebra which satisfy
\begin{equation}
\left[S_{\hat\mu\hat\nu},\,S_{\hat\sigma\hat\tau}\right]=i(
\eta_{\hat\mu\hat\tau}\,S_{\hat\nu\hat\sigma}-
\eta_{\hat\mu\hat\sigma}\,S_{\hat\nu\hat\tau}+
\eta_{\hat\nu\hat\sigma}\,S_{\hat\mu\hat\tau}-
\eta_{\hat\nu\hat\tau}\,S_{\hat\mu\hat\sigma})\,.
\end{equation}
For small values of $\omega^{\hat\alpha\hat\beta}$ the matrix elements of the
transformations $\Lambda$ can be written as
\begin{equation}\label{infLam}
\Lambda[A(\omega)]^{\hat\mu\,\cdot}_{\cdot\,\hat\nu}=\delta^{\hat\mu}_{\hat\nu}
+\omega^{\hat\mu\,\cdot}_{\cdot\,\hat\nu}
+\cdots\,.
\end{equation}

Now we assume that $M$ is orientable and time-orientable such that
$L^{\uparrow}_{+}$ can be considered as the gauge group of the Minkowski
metric \cite{WALD}. Then the  fields with spin can be defined as in the case
of the flat spacetime, with the help of the finite-dimensional {\em linear}
representations, $\rho$, of the $SL(2,\Bbb C)$ group \cite{W}. In general, the
fields $\psi_{\rho}:M\to V_{\rho}$ are defined over $M$ with values in the
vector spaces $V_{\rho}$ of the representations  $\rho$. In the following we
systematically use the bases of $V_{\rho}$ labeled only by spinor or vector
{\em local} indices defined with respect to the axes of the local frames
given by the tetrad fields. These will not be written explicitly except the
cases when this is requested by the concrete calculation needs.

The relativistic covariant field equations are derived from actions
\cite{UK,SW},
\begin{equation}\label{act}
{\cal S}[\psi_{\rho},e]=\int d^{4}x\sqrt{g}\,{\cal L}(
\psi_{\rho}, D_{\hat\mu}^{\rho}\psi_{\rho})\,,
\quad g=|\det(g_{\mu\nu})|\,,
\end{equation}
depending on the matter fields, $\psi_{\rho}$, and the components
of the tetrad fields, $e$, which represent the gravitational
degrees of freedom. Recently it was shown that this action can be
completed adding a term with the integration measure $d^{4}x\Phi$
(instead of $d^{4}x \sqrt{g}$) where $\Phi$ can be expressed in
terms of scalar fields independent on $e$ \cite{GUEN}. This new
term allows one to define a new global scale symmetry which, in
our opinion, is compatible with the geometric symmetries we study
here. Therefore, without to lose generality, we can restrict
ourselves to actions of the traditional form (\ref{act}) in which
the canonical variables are the components of the fields
$\psi_{\rho}$ and $e$.

The  covariant derivatives,
\begin{equation}\label{covder}
D_{\hat\alpha}^{\rho}=\hat e_{\hat\alpha}^{\mu}D_{\mu}^{\rho}=
\hat\partial_{\hat\alpha}+\frac{i}{2}\,
\rho(S^{\hat\beta\, \cdot}
_{\cdot\, \hat\gamma})\,\hat\Gamma^{\hat\gamma}_{\hat\alpha \hat\beta}\,,
\end{equation}
assure  the invariance of the whole theory under the {\em tetrad gauge}
transformations,
\begin{eqnarray}
\hat e^{\hat\alpha}_{\mu}(x)&\to& \hat e'^{\hat\alpha}_{\mu}(x)=
\Lambda[A(x)]^{\hat\alpha\,\cdot}_{\cdot\,\hat\beta}
\,\hat e^{\hat\beta}_{\mu}(x)\,,\nonumber\\
e_{\hat\alpha}^{\mu}(x)&\to&  {e'}_{\hat\alpha}^{\mu}(x)=
\Lambda[A(x)]_{\hat\alpha\,\cdot}^{\cdot\,\hat\beta}
\,e_{\hat\beta}^{\mu}(x)\,,\label{gauge}\\
\psi_{\rho}(x)&\to&~\psi_{\rho}'(x)=\rho[A(x)]\,\psi_{\rho}(x)\,,\nonumber
\end{eqnarray}
determined by the mappings $A:M\to SL(2,\Bbb C)$ the values of which are the
local $SL(2,\Bbb C)$ transformations $A(x)\equiv A[\omega(x)]$. These mappings can
be organized as a group, ${\cal G}$, with respect to the multiplication
$~\times ~$ defined as $(A'\times A)(x)=A'(x)A(x)$. The notation
$Id$ stands for the mapping identity, $Id(x)=1\in SL(2,\Bbb C)$, while $A^{-1}$
is the inverse of $A$, $(A^{-1})(x)=[A(x)]^{-1}$.

\subsection{Combined transformations}

The general coordinate transformations are automorphisms of $M$
which, in the passive mode, can be seen as changes of the local
charts corresponding to the same domain of $M$ \cite{WALD,ON}. If
$x$ and $x'$ are the coordinates of a  point in two different
charts then there is a mapping $\phi$ between these charts giving
the coordinate transformation, $x\to x'=\phi(x)$. These
transformations  form a group with respect to the composition of
mappings, $\,\circ\,$, defined as usual, i.e.
$(\phi'\circ\phi)(x)=\phi'[\phi(x)]$. We denote this group by
${\cal A}$, its identity map by $id$ and the inverse mapping of
$\phi$ by $\phi^{-1}$.

The automorphisms change all the components carrying  natural indices
including those of the tetrad fields \cite{SW} changing thus the positions of
the local frames with respect to the natural ones. If we assume that the
physical experiment makes reference to the axes of the local frame then it
could appear situations when several correction of the positions of these
frames  should be needed before (or after) a general coordinate
transformation. Obviously, these have to be done with the help of suitable
gauge transformation  associated to the automorphisms. Thus we arrive to the
necessity of introducing  the {\em combined} transformations denoted by
$(A,\phi)$ and defined as gauge transformations, given by $A\in {\cal G}$,
followed by automorphisms, $\phi\in{\cal A}$. In this new notation the pure
gauge transformations will appear as $(A,id)$ while the automorphisms will be
denoted from now by $(Id,\phi)$.

The  effect of a combined transformation $(A,\phi)$ upon our basic fields,
$\psi_{\rho},\, e$ and $\hat e$ is
$x\to x'=\phi(x),\,e(x)\to
e'(x'),\, \hat e(x)\to \hat e'(x')$ and $\psi_{\rho}(x)\to \psi'_{\rho}(x')
=\rho[A(x)]\psi_{\rho}(x)$ where $e'$ are the transformed tetrads of the
components
\begin{equation}
e'^{\mu}_{\hat\alpha}[\phi(x)]=\Lambda[A(x)]^{\cdot\,\hat\beta}
_{\hat\alpha\,\cdot}\,e^{\nu}_{\hat\beta}(x)
\frac{\partial\phi^{\mu}(x)}{\partial x^{\nu}}\,,
\end{equation}
while the components of $\hat e'$ have to be calculated according to
Eqs.(\ref{(duale)}). Thus we have written down the most general
transformation laws that leave  the action invariant in the sense
that ${\cal S}[\psi_{\rho}',e']={\cal S}[\psi_{\rho},e]$. The field
equations derived from ${\cal S}$, written in local frames as
$(E_{\rho}\psi_{\rho})(x)=0$, {\em covariantly} transform according
to the rule
\begin{equation}\label{ero}
(E_{\rho}\psi_{\rho})(x) \to
(E'_{\rho}\psi'_{\rho})(x')=\rho[A(x)](E_{\rho}\psi_{\rho})(x)\,,
\end{equation}
since the operators $E_{\rho}$ involve covariant derivatives \cite{SW}.

The association among the transformations of the groups ${\cal G}$ and
${\cal A}$ must lead to a new group with a specific multiplication.  In order
to find how looks this new operation  it is convenient to use the
composition among  the  mappings $A$ and $\phi$ (taken only in this order)
giving new mappings, $A\circ\phi\in {\cal G}$, defined as $(A\circ \phi)(x)=A[\phi(x)]$.
The  calculation rules  $Id\circ \phi=Id$, $A\circ id=A$ and
$(A'\times A)\circ \phi=(A'\circ \phi)\times (A\circ \phi)$ are obvious.
With these ingredients we define the new multiplication
\begin{equation}\label{comp}
(A',\phi')*(A,\phi)=
\left((A'\circ\phi)\times A,\phi'\circ\phi\right)\,.
\end{equation}
It is clear that now the identity is $(Id,id)$  while the inverse of a pair
$(A,\phi)$ reads
\begin{equation}\label{compAphi}
(A,\phi)^{-1}=(A^{-1}\circ\phi^{-1},\phi^{-1})\,.
\end{equation}
First of all we observe that the operation $~*~$ is well-defined and
represents
the composition among the combined transformations since these can be
expressed, according to their definition, as  $(A,\phi)=(Id,\phi)*(A,id)$.
Furthermore,  we can convince ourselves that if we perform successively two
arbitrary combined transformations, $(A,\phi)$ and $(A',\phi')$, then
the resulting transformation is just $(A',\phi')*(A,\phi)$ as given by
Eq.(\ref{comp}). This means that the combined transformations form a group
with respect to the multiplication $\,*\,$.
It is not difficult to verify that this group, denoted by $\tilde{\cal G}$,
is the semidirect product $\tilde{\cal G}={\cal G}\circledS{\cal A}$ where
${\cal G}$ is the {\em invariant} subgroup while ${\cal A}$ is an usual one.

In the theories involving only vector and tensor fields we do not need to use
the combined transformations  defined above since the theory is independent
on the positions of the local frames. This can be easily shown even in our
approach where we use field components with local indices. Indeed, if we
perform a  combined transformation $(A,\phi)$ then  any tensor field  of
rank  $(p,q)$,
\begin{equation}
\psi^{\hat\alpha_{1}, \hat\alpha_{2},...,\hat\alpha_{p}}
_{\hat\beta_{1}, \hat\beta_{2},...,\hat\beta_{q}}=
\hat e^{\hat\alpha_{1}}_{\mu_{1}}\cdots
\hat e^{\hat\alpha_{p}}_{\mu_{p}}\,
 e_{\hat\beta_{1}}^{\nu_{1}}\cdots
 e_{\hat\beta_{q}}^{\nu_{q}}\,
\psi^{\mu_{1},\mu_{2},...,\mu_{p}}
_{\nu_{1},\nu_{2},...,\nu_{q}}\,,
\end{equation}
transforms according to the representation
\begin{equation}
\rho_{\hat\alpha_{1}, \hat\alpha_{2},...,\hat\alpha_{p};\,
\hat\beta'_{1}, \hat\beta'_{2},...,\hat\beta'_{q}}
^{\hat\beta_{1}, \hat\beta_{2},...,\hat\beta_{q};\,
\hat\alpha'_{1}, \hat\alpha'_{2},...,\hat\alpha'_{p}}(A)=
\Lambda^{\hat\beta_{1}\,\cdot}_{\cdot\,\hat\beta'_{1}}(A)\cdots\,
\Lambda^{\cdot\, \hat\alpha'_{1}}_{\hat\alpha_{1}\,\cdot}(A)\cdots\,,
\end{equation}
such that the resulting transformation law of the components carrying natural
indices,
\begin{equation}
\psi'^{\mu_{1},...}
_{\,\nu_{1},...}(x')=
\frac{\partial x'^{\mu_{1}}}{\partial x^{\sigma_{1}}}\cdots\,
\frac{\partial x^{\tau_{1}}}{\partial x'^{\nu_{1}}}\cdots\,
\psi^{\sigma_{1},....}_{\tau_{1},....}(x)\,,
\end{equation}
is just the familiar one \cite{SW}. In other words, in this case the effect
of the combined transformations reduces to that of their automorphisms.
However, when the half integer spin fields are  involved  this is no
more true and we must use the combined transformations of  $\tilde{\cal G}$
if we want to keep under control the positions of the local frames.

\section{External symmetry}

In general, the symmetry of any manifold $M$ is given by its
isometry group whose transformations leave invariant the metric
tensor in any chart. The scalar field transforms under isometries
according to the standard scalar representation generated by the
orbital generators related to the Killing vectors of $M$
\cite{SW,WALD,ON}. In the following we present the generalization
of this theory of symmetry to fields with spin, for which we have
defined the {\em external} symmetry group and its representations
\cite{C2}.

\subsection{Isometries}

There are conjectures when several
coordinate transformations, $x\to x'=\phi_{\xi}(x)$, depend on $N$
independent real parameters, $\xi^a$ ($a,b,c...=1,2,...,N$),
such that $\xi=0$ corresponds to the identity map, $\phi_{\xi=0}=id$.
The set of these mappings  is a Lie group \cite{GIL},
$G\in {\cal G}$, if they accomplish the composition rule
\begin{equation}\label{compphi}
\phi_{\xi'}\circ \phi_{\xi}=\phi_{f(\xi',\xi)}\,,
\end{equation}
where the functions $f: G\times G\to G$  define the group multiplication.
These must satisfy $f^{a}(0,\xi)=f^{a}(\xi,0)=\xi^{a}$ and
 $f^{a}(\xi^{-1},\xi)=f^{a}(\xi,\xi^{-1})=0$ where $\xi^{-1}$ are the
parameters of the inverse mapping of $\phi_{\xi}$,
$\phi_{\xi^{-1}}=\phi^{-1}_{\xi}$.
Moreover, the structure constants of $G$ can be calculated as \cite{HAM}
\begin{equation}\label{c}
c_{abc}=\left(\frac{\partial f^{c}(\xi,\xi')}{\partial \xi^{a}\partial \xi'^{b}}-
\frac{\partial f^{c}(\xi,\xi')}{\partial \xi^{b}\partial \xi'^{a}}
\right)_{|\xi=\xi'=0}\,.
\end{equation}
For small values of the group parameters the infinitesimal transformations,
$x^{\mu}\to x'^{\mu}=x^{\mu}+\xi^{a}k_{a}^{\mu}(x)+\cdots$,
are given by the vectors $k_{a}$ whose components,
\begin{equation}\label{ka}
k_{a}^{\mu}={\frac{\partial \phi^{\mu}_{\xi}}{\partial\xi^{a}}}_{|\xi=0}\,,
\end{equation}
satisfy the  identities
\begin{equation}\label{kkc}
k^{\mu}_{a}k^{\nu}_{b,\mu}
-k^{\mu}_{b}k^{\nu}_{a,\mu}+c_{abc}k^{\nu}_{c}=0\,,
\end{equation}
resulting from Eqs.(\ref{compphi}) and (\ref{c}).

In the following we restrict ourselves to consider only the {\em isometry}
transformations, $x'=\phi_{\xi}(x)$, which leave invariant the components of
the metric tensor \cite{SW,ON}, i.e.
\begin{equation}\label{giso}
g_{\alpha\beta}(x')\frac{\partial x'^\alpha}{\partial x^\mu}
\frac{\partial x'^\beta}{\partial x^\nu}=g_{\mu\nu}(x)\,.
\end{equation}
These form the isometry group $G\equiv I(M)$ which is the Lie group giving
the symmetry of the spacetime $M$. We consider that this has $N$ independent
parameters and, therefore,  $k_{a}, a=1,2,...N$, are independent Killing
vectors (which satisfy $k_{a\, \mu;\nu}+k_{a\, \nu;\mu}=0$). Then their
corresponding Lie derivatives form a basis of the Lie algebra $i(M)$ of the
group $I(M)$ \cite{ON}.

However, in practice we are interested to find the operators of the
relativistic quantum
theory related to these geometric objects which describe the symmetry of
the background. For this reason we focus upon the operator-valued
representations \cite{BR} of the group $I(M)$ and its algebra. The
{\em scalar} field $\psi:M\to \Bbb C$ transforms under isometries as
$\psi(x)\to \psi'[\phi_{\xi}(x)]=\psi(x)$. This rule defines the
representation $\phi_{\xi}\to T_{\xi}$ of the group $I(M)$ whose
operators have the action $\psi'=T_{\xi}\psi=\psi\circ\phi^{-1}_{\xi}$.
Hereby it results that the operators of
infinitesimal transformations, $T_{\xi}=1-i\xi^{a}L_{a}+\cdots$, depend on
the basis-generators,
\begin{equation}\label{La}
L_{a}=-ik_{a}^{\mu}\partial_{\mu}\,, \quad a=1,2,...,N\,,
\end{equation}
which are completely determined by the Killing vectors. From Eq.(\ref{kkc})
we see that they obey the commutation rules
\begin{equation}\label{comL}
[L_{a}, L_{b}]=ic_{abc}L_{c}\,,
\end{equation}
given by the structure constants of $I(M)$. In other words they form a basis
of the operator-valued representation of the Lie algebra $i(M)$ in a carrier
space of scalar fields. Notice that in the usual quantum mechanics the
operators similar to the generators $L_{a}$ are called often {\em orbital}
generators.

\subsection{The group of external symmetry}

Now  the problem is  how may transform  under isometries the whole
geometric framework of the theories with spin where we explicitly
use the local frames. Since the isometry is a general coordinate
transformation it changes the relative positions of the local and
natural frames. This fact may be an impediment when one intends to
study the symmetries of the theories with spin induced by  those
of the background. For this reason it is natural to suppose that
the good symmetry  transformations we need are combined
transformations in which the isometries are preceded by
appropriate gauge transformations such that not only  the form of
the metric tensor should be conserved but the form of the tetrad
field components too.

Thus we arrive at the main point of our theory. We introduce the
{\em external symmetry} transformations, $(A_{\xi},\phi_{\xi})$,
as combined transformations involving isometries and corresponding
gauge transformations necessary to {\em preserve the gauge}. We
assume that in a fixed gauge, given by the tetrad fields $e$ and
$\hat e$, $A_{\xi}$ is defined by
\begin{equation}\label{Axx}
\Lambda[A_{\xi}(x)]^{\hat\alpha\,\cdot}_{\cdot\,\hat\beta}=
\hat e_{\mu}^{\hat\alpha}[\phi_{\xi}(x)]\frac{\partial \phi^{\mu}_{\xi}(x)}
{\partial x^{\nu}}\,e^{\nu}_{\hat\beta}(x)\,,
\end{equation}
with the supplementary condition $A_{\xi=0}(x)=1\in SL(2,\Bbb C)$.
Since $\phi_{\xi}$ is an isometry Eq.(\ref{giso}) guarantees that
$\Lambda[A_{\xi}(x)]\in L^{\uparrow}_{+}$ and, implicitly,
$A_{\xi}(x)\in SL(2,\Bbb C)$. Then the transformation laws of our fields are
\begin{equation}\label{es}
(A_{\xi},\phi_{\xi}):\qquad
\begin{array}{rlrcl}
x&\to&x'&=&\phi_{\xi}(x)\,,\\
e(x)&\to&e'(x')&=&e[\phi_{\xi}(x)]\,,\\
\hat e(x)&\to&\hat e'(x')&=&\hat e[\phi_{\xi}(x)]\,,\\
\psi_{\rho}(x)&\to&\psi_{\rho}'(x')&=&\rho[A_{\xi}(x)]\psi_{\rho}(x)\,.
\end{array}
\qquad
\end{equation}
The mean virtue of these transformations is that they leave {\em invariant}
the form of the operators of the field equations, $E_{\rho}$, in local frames.
This is because the components of the tetrad fields and, consequently, the
covariant derivatives in local frames, $D^{\rho}_{\hat\mu}$, do not change
their form.

For small $\xi^{a}$ the covariant $SL(2,\Bbb C)$  parameters
of $A_{\xi}(x)\equiv A[\omega_{\xi}(x)]$ can be written as
$\omega^{\hat\alpha\hat\beta}_{\xi}(x)=
\xi^{a}\Omega^{\hat\alpha\hat\beta}_{a}(x)+\cdots$ where,
according to  Eqs.(\ref{Aomega}), (\ref{infLam})
and (\ref{Axx}), we have
\begin{equation}\label{Om}
\Omega^{\hat\alpha\hat\beta}_{a}\equiv {\frac{\partial
\omega^{\hat\alpha\hat\beta}_{\xi}}
{\partial\xi^a}}_{|\xi=0}
=\left( \hat e^{\hat\alpha}_{\mu}\,k_{a,\nu}^{\mu}
+\hat e^{\hat\alpha}_{\nu,\mu}
k_{a}^{\mu}\right)e^{\nu}_{\hat\lambda}\eta^{\hat\lambda\hat\beta}\,.
\end{equation}
We must specify that these functions are antisymmetric if and only if $k_{a}$
are Killing vectors. This indicates that the association among isometries
and the gauge transformations defined by Eq.(\ref{Axx}) is correct.

It remains to show that the transformations
$(A_{\xi},\phi_{\xi})$ form a Lie group related to
$I(M)$. Starting with Eq.(\ref{Axx}) we find that
\begin{equation}\label{compA}
(A_{\xi'}\circ\phi_{\xi})\times A_{\xi}=A_{f(\xi',\xi)}\,,
\end{equation}
and, according to Eqs.(\ref{comp}) and (\ref{compphi}), we obtain
\begin{equation}\label{mult}
(A_{\xi'},\phi_{\xi'})*(A_{\xi},\phi_{\xi})=
(A_{f(\xi',\xi)},\phi_{f(\xi',\xi)})\,,
\end{equation}
and $(A_{\xi=0},\phi_{\xi=0})=(Id,id)$.
Thus we have shown that the  pairs $(A_{\xi},\phi_{\xi})$
form a Lie group with respect to the operation $~*~$.
We say that this is the  external symmetry group of $M$ and we denote it
by $S(M)\subset \tilde{\cal G}$. From Eq.(\ref{mult}) we understand that $S(M)$
is {\em locally
isomorphic} with $I(M)$ and, therefore, the Lie algebra of $S(M)$, denoted by
$s(M)$, is isomorphic with $i(M)$ having the same structure constants.
In our opinion, $S(M)$ must be isomorphic with the universal covering group of
$I(M)$ since it has  anyway the topology induced by $SL(2,\Bbb C)$ which is simply
connected.
In general, the number of group parameters of $I(M)$ or $S(M)$
(which is equal to the number of the independent Killing vectors of $M$)
can be $0\le N\le 10$.

The form of the external symmetry transformations is strongly dependent on
the choice of the local chart as well as that of the tetrad gauge. If
we change simultaneously the gauge and the coordinates with the help of a
combined transformation $(A,\phi)$ then each
$(A_{\xi},\phi_{\xi})\in S(M)$ transforms as
\begin{equation}\label{aaprim}
(A_{\xi},\phi_{\xi})\to
(A'_{\xi},\phi'_{\xi})=(A,\phi)*(A_{\xi},\phi_{\xi})*(A,\phi)^{-1}
\end{equation}
which means that
\begin{eqnarray}\label{apsiprim}
A'_{\xi}&=&\left\{\left[\left(A\circ\phi_{\xi}\right)\times A_{\xi}\right]
\times A^{-1}\right\}\circ \phi^{-1}\,,\\
\phi'_{\xi}&=&\left(\phi\circ\phi_{\xi}\right)\circ\phi^{-1}\,.
\end{eqnarray}
Obviously, these transformations define automorphisms of $S(M)$.

\subsection{Representations}

The last of Eqs.(\ref{es}) which gives the transformation law of
the field $\psi_{\rho}$ defines the operator-valued representation
$(A_{\xi},\phi_{\xi})\to T_{\xi}^{\rho}$ of the group $S(M)$,
\begin{equation}\label{rep}
(T_{\xi}^{\rho}\psi_{\rho})[\phi_{\xi}(x)]=\rho[A_{\xi}(x)]\psi_{\rho}(x)\,.
\end{equation}
The mentioned invariance under these transformations of the operators of the
field equations in local frames reads
\begin{equation}\label{invE}
T^{\rho}_{\xi}E_{\rho}{(T^{\rho}_{\xi})}^{-1}=E_{\rho}\,.
\end{equation}
Since  $A_{\xi}(x)\in SL(2,\Bbb C)$  we  say that this representation is
{\em induced} by the representation $\rho$ of $SL(2,\Bbb C)$ \cite{BR,MAK}.
As we have shown in Sec.2.2, if $\rho$ is a vector or tensor representation
(having only integer spin components) then the effect of the transformation
(\ref{rep}) upon the components carrying natural indices is due only to
$\phi_{\xi}$. However, for the representations with half integer spin the
presence of $A_{\xi}$ is crucial since there are no natural indices.
In addition, this allows us to define  the generators of the
representations (\ref{rep}) for any spin.

The basis-generators of the representations of the Lie algebra
$s(M)$ are the operators
\begin{equation}\label{X}
X^{\rho}_{a} = i{\frac{\partial T_{\xi}^{\rho}}{\partial \xi^{a}}}_{|\xi=0}=
L_{a}+S_{a}^{\rho}\,,
\end{equation}
which appear as  sums among the orbital generators defined by
Eq.(\ref{La}) and the {\em spin terms}  which have the action
\begin{equation}\label{sss}
(S_{a}^{\rho}\psi_{\rho})(x)=\rho[S_{a}(x)]\psi_{\rho}(x)\,.
\end{equation}
This is determined by the form of the {\em local} $SL(2,\Bbb C)$ generators,
\begin{equation}\label{Sx}
S_{a}(x)=i{\frac{\partial A_{\xi}(x)}{\partial \xi^{a}}}_{|\xi=0}=
\frac{1}{2}\Omega^{\hat\alpha\hat\beta}_{a}(x)
S_{\hat\alpha\hat\beta}\,,
\end{equation}
that depend on the functions (\ref{Om}).
Furthermore,  if we derive  Eq.(\ref{compA}) with respect to $\xi$ and $\xi'$
then from Eqs.(\ref{infLam}), (\ref{c}) and (\ref{Om}), after a few
manipulations, we obtain the identities
\begin{equation}\label{idOM}
\eta_{\hat\alpha\hat\beta}\left(
\Omega_{a}^{\hat\alpha\hat\mu}\Omega_{b}^{\hat\beta\hat\nu}
-\Omega_{b}^{\hat\alpha\hat\mu}\Omega_{a}^{\hat\beta\hat\nu}\right)+
k^{\mu}_{a}\Omega_{b,\mu}^{\hat\mu\hat\nu}
-k^{\mu}_{b}\Omega_{a,\mu}^{\hat\mu\hat\nu}+c_{abc}\Omega_{c}^{\hat\mu\hat\nu}
=0\,.
\end{equation}
Hereby it results that
\begin{equation}
[S_{a}^{\rho},S_{b}^{\rho}]+[L_{a},S_{b}^{\rho}]-[L_{b},S_{a}^{\rho}]
=ic_{abc}S_{c}^{\rho}\,,
\end{equation}
and, according to Eq.(\ref{comL}), we find the expected commutation rules
\begin{equation}\label{comX}
[X_{a}^{\rho}, X_{b}^{\rho}]=ic_{abc}X_{c}^{\rho}\,.
\end{equation}
Thus we have derived the basis-generators of the operator-valued representation
of $s(M)$  induced by the linear representation $\rho$ of $SL(2,\Bbb C)$. All
the operators of this representation commute with the operator $E_{\rho}$
since, according to Eqs.(\ref{invE}) and (\ref{X}), we have
\begin{equation}
[E_{\rho},X^{\rho}_{a}]=0\,, \quad a=1,2,...,N\,.
\end{equation}
Therefore, for defining quantum modes we can use the  set of commuting
operators containing  the Casimir operators of $s(M)$, the operators of its
Cartan subalgebra and $E_{\rho}$.

Finally, we must specify that the basis-generators (\ref{X}) of
the representations of the $s(M)$ algebra can be written in
covariant form as
\begin{equation}
X^{\rho}_{a}=-ik^{\mu}_{a}D_{\mu}^{\rho}+\frac{1}{2}\,
k_{a\, \mu;\nu}\,e^{\mu}_{\hat\alpha}\,e^{\nu}_{\hat\beta}\,
\rho(S^{\hat\alpha\hat\beta})\,,
\end{equation}
generalizing thus the important result obtained in  Ref.
\cite{CART} for the Dirac field.

\subsection{Manifest covariance}

The action of the operators $X_{a}^{\rho}$ depends on the choice of
many elements: the natural coordinates, the tetrad gauge, the group
parametrization and the representation $\rho$. What is important
here is that they are strongly dependent  on the tetrad gauge fixing
even in the case of the representations with integer spin. This is
because the covariant parametrization of the $SL(2,\Bbb C)$ algebra
is defined with respect to the axes of the local frames.  In
general, if we consider the representation $(A_{\xi},\phi_{\xi})\to
T^{\rho}_{\xi}$ and we perform the transformation (\ref{aaprim})
then it results the {\em equivalent} representation,
$(A'_{\xi},\phi'_{\xi})\to T'^{\rho}_{\xi}$. Its generators
calculated  from Eqs.(\ref{apsiprim}) indicate that in this case the
equivalence relations are much more complicated than those of the
usual theory of linear representations. Without to enter in other
technical details we specify that if we change only the gauge with
the help of the transformation $(A, id)$ then the local $SL(2,\Bbb
C)$ generators (\ref{Sx}) transform as
\begin{eqnarray}
S_{a}(x)\to S'_{a}(x)&=& A(x)S_{a}(x) A(x)^{-1}\nonumber\\
&&+k_{a}^{\sigma}(x)\Lambda[A(x)]_{\hat\alpha
\hat\mu,\,\sigma}\Lambda[A(x)]_{\hat\beta\,\cdot}^{\cdot\,\hat\mu}
S^{\hat\alpha\hat\beta}\,,
\end{eqnarray}
while the orbital parts do not change their form. This means that the gauge
transformations change, in addition, the commutation relations among the spin
and orbital parts of the generators $X_{a}^{\rho}$.

The consequence is that we can find gauge fixings where the local $SL(2,\Bbb C)$
generators $S_{a}(x)$, $a=1,2,...,n$ ($n\le N$), corresponding to a subgroup
$H\subset S(M)$, are independent on $x$ and, therefore,
$[S_{a}^{\rho},L_{b}]=0$ for all $a=1,2,...,n$  and $b=1,2,...,N$. Then the
operators $S_{a}^{\rho},\, a=1,2,...,n$  are just the basis-generators of an
usual linear representation of $H$  and the field $\psi_{\rho}$
behaves {\em manifestly covariant} under the external symmetry
transformations of this subgroup. Of course, when $H=S(M)$  we say simply
that the field $\psi_{\rho}$ is manifest covariant.

The simplest examples are the manifest covariant fields of special relativity.
Since here the spacetime $M$ is flat, the metric in Cartesian coordinates is
$g_{\mu\nu}=\eta_{\mu\nu}$ and one can use the {\em inertial} (local) frames
with $e^{\mu}_{\nu}=\hat e^{\mu}_{\nu}=\delta^{\mu}_{\nu}$. Then
the isometries are just the transformations $x'=\Lambda[A(\omega)]x -a$ of
the Poincar\' e group,
${\cal P}_{+}^{\uparrow} = T(4)\circledS L_{+}^{\uparrow}$  \cite{W}.
If we denote by $\xi^{(\mu\nu)}=\omega^{\mu\nu}$ the $SL(2,\Bbb C)$ parameters and
by $\xi^{(\mu)}=a^{\mu}$ those of the translation group $T(4)$, then
it is a simple exercise to calculate the basis-generators
\begin{eqnarray}
X_{(\mu)}^{\rho}&=&i\partial_{\mu}\,,\\
X_{(\mu\nu)}^{\rho}&=&i(\eta_{\mu\alpha}x^{\alpha}
\partial_{\nu}- \eta_{\nu\alpha}x^{\alpha}\partial_{\mu})+
\rho(S_{\mu\nu})\,,
\end{eqnarray}
which show us that $\psi_{\rho}$ transforms manifestly covariant.
On the other hand, it is clear that the group
$S(M)\equiv \tilde{\cal P}^{\uparrow}_{+}=T(4)\circledS SL(2,\Bbb C)$
is just the universal covering group of $I(M)\equiv{\cal P}_{+}^{\uparrow}$.

In general, there are many cases of curved spacetimes for which one can choose
suitable local frames allowing one to introduce manifest covariant fields with
respect to a subgroup $H\subset S(M)$ or even  the whole group $S(M)$.
In our opinion, this is possible only when $H$ or $S(M)$ are at most subgroups
of $\tilde{\cal P}_{+}^{\uparrow}$.

\section{The central symmetry}

Let us take as first example the spacetimes $M$ which have
spherically symmetric static chart that will be called here {\em
central} charts (or central frames). These manifolds have the
isometry group $I(M)=T(1)\otimes SO(3)$ of time translations and
space rotations.

\subsection{Central charts}

In a central chart $\{t,\vec{x} \}$, with Cartesian coordinates
$x^{0}=t$ and  $x^i$ ($i,j,k...=1,2,3$),  the metric tensor is
time-independent and transforms manifestly covariant  under the
rotations $R\in SO(3)$ of the space coordinates,
\begin{equation}\label{(rot)}
t'=t,\quad  x'^{i}= R_{\cdot\,j}^{i\,\cdot}(\omega)x^{j}=x^{i}+
\omega^{i\,\cdot}_{\cdot\,j}x^{j}+\cdots\,,
\end{equation}
denoted simply by $x\to x'=Rx$. Here the most general
form of the line element,
\begin{equation}\label{(metr)}
ds^{2}=g_{\mu\nu}(x)dx^{\mu}dx^{\nu}=A(r)dt^{2}-[B(r)\delta_{ij}+
C(r)x^{i}x^{j}]dx^{i}dx^{j}\,,
\end{equation}
may involve three functions, $A$, $B$ and $C$, depending only on the
Euclidean norm of $\vec{x}$, $r=|\vec{x}|$. In applications it is
convenient to replace these functions by  new ones,  $u$,  $v$ and
$w$, defined as
\begin{equation}\label{(ABC)}
A=w^{2}, \quad B=\frac{w^2}{v^2}, \quad
C=\frac{1}{r^2}\left( \frac{w^2}{u^2}-\frac{w^2}{v^2}\right)\,.
\end{equation}
Therefore, in this chart we have
\begin{equation}\label{ggg}
\sqrt{g}=B[A(B+r^{2}C)]^{1/2}=\frac{w^4}{uv^2}\,.
\end{equation}

Other useful central charts are those with  spherical coordinates,
$\{t,r,\theta,\phi\}$, commonly associated with the Cartesian space
ones, $x^i$. In these charts the line elements have the form
\begin{equation}\label{(muvw)}
ds^{2}=g_{\mu\nu}^s(x)dx^{\mu}dx^{\nu}=w^{2}dt^{2}-\frac{w^2}{u^2}dr^{2}-
\frac{w^2}{v^2}r^{2}(d\theta^{2}+\sin^{2}\theta d\phi^{2})\,,
\end{equation}
and $\sqrt{g^s}=\sqrt{g}\,r^2\sin(\theta)$.

The new functions we introduced here have simple transformation
laws under the isotropic dilatations which change only the radial
coordinate, $r\to r'(r)$, without to affect the central symmetry
of the line element. These transformations,
\begin{equation}
u'(r')=u(r)\left|\frac{dr'(r)}{dr}\right|\,,\quad
v'(r')=v(r)\frac{r'(r)}{r}\,,\quad
w'(r')=w(r)\,,
\end{equation}
allow one to choose desired forms for the functions $u,v$ and $w$.

\subsection{The Cartesian gauge}

The Cartesian gauge in central charts was  mentioned  long time
ago \cite{BW} but it is less used in concrete problems since it
leads to complicated calculations in spherical coordinates.
However, in Cartesian coordinates this gauge has the advantage of
explicitly pointing out the global central symmetry of the
manifold. In Refs. \cite{COTA} we have proposed a version of
Cartesian gauge in central charts with Cartesian coordinates that
preserve the manifest covariance under rotations (\ref{(rot)}) in
the sense that the  1-forms $d\hat x^{\hat\mu}=\hat
e^{\hat\mu}_{\alpha}(x)dx^{\alpha}$ transform as
\begin{equation}\label{(tr)}
d\hat x ^{\hat\mu}\to d\hat x'^{\hat\mu}=\hat e^{\hat\mu}_{\alpha}(x')
dx'^{\alpha}=(Rd\hat x)^{\hat\mu}.
\end{equation}
If the line element has the form (\ref{(metr)}) then a good choice
of the tetrad fields with the above property is
\begin{eqnarray}
&&\hat e^{0}_{0}=\hat a(r), \quad \hat e^{0}_{i}=\hat e^{i}_{0}=0, \quad
\hat e^{i}_{j}=\hat b(r)\delta_{ij}+\hat c(r) x^{i}x^{j}
\,\label{(eee)}\\
&&e^{0}_{0}= a(r), \quad  e^{0}_{i}= e^{i}_{0}=0, \quad e^{i}_{j}=
b(r)\delta_{ij}+ c(r) x^{i}x^{j} \,,\label{(eee1)}
\end{eqnarray}
where, according to Eqs. (\ref{gmunu}), (\ref{(metr)}) and
(\ref{(ABC)}), we must have
\begin{eqnarray}
&&\hat a=w\,,\quad  \hat b=\frac{w}{v}\,,\quad \hat
c=\frac{1}{r^2} \left(
\frac{w}{u}-\frac{w}{v}\right)\,, \label{(abc)}\\
&&a= \frac{1}{w}\,, \quad  b=\frac{v}{w}\,, \quad c=\frac{1}{r^2}
\left( \frac{u}{w}-\frac{v}{w}\right)\,. \label{(abc1)}
\end{eqnarray}

When one defines the metric tensor such that
{${g_{\mu\nu}}_{|r=0}=\eta_{\mu\nu}$ then
$u(0)^{2}=v(0)^{2}=w(0)^{2}=1$.  In other respects, from Eqs.
(\ref{(abc)}) and (\ref{(abc1)}) we see that the function $w$ must
be positively defined in order to keep the same sense for the time
axes of the natural and local frames. In addition, it is convenient
to consider that the function $u$ is positively defined too.
However, the function $v=\eta_{P}|v|$ has the sign given by the
relative parity $\eta_{P}$ which takes the value  $\eta_{P}=1$ when
the space axes of the local frame at $x=0$ are parallel with those
of the natural frame, and $\eta_{P}=-1$ if these are anti-parallel.

Now we have all the elements we need to calculate the generators
of the representations $T^{\rho}$ of the group $S(M)$. If we
denote by $\xi^{(0)}$ the parameter of the time translations and
by $\xi^{(i)}=\varepsilon_{ijk}\omega^{jk}/2$ the parameters of
the rotations (\ref{(rot)}), we find  that the local $SL(2,\Bbb
C)$ generators of Eq. (\ref{Sx}) are just the $su(2)$ ones, i.e.
$S_{(i)}(x)=S_{i}= \varepsilon_{ijk}S_{jk}/2$, such that the
basis-generators read
\begin{equation}\label{ang1}
X_{(0)}^{\rho}\equiv H=i\partial_{t}\,,\quad X_{(i)}^{\rho} \equiv
J_{(i)}^{\rho}=L_{(i)}+\rho(S_{i})
\end{equation}
where $L_{(i)}=-i\varepsilon_{ijk}x^{j}\partial_{k}$ are the usual
components of the orbital angular momentum. Thus we obtain that the
group $S(M)=T(1)\otimes SU(2)$ is the universal covering group of
$I(M)$. Its transformations are gauge transformations
$A_{\vec{\xi}}\in SU(2)$, independent on $x$, combined with the
isometries of $I(M)$ given by  $x\to x'=R(A_{\vec{\xi}})\,x$ and
$t\to t'=t-\xi^{(0)}$. This means that, in this gauge, the field
$\psi_{\rho}$  transforms manifestly covariant. The generators have
the usual physical significance, namely $H$ is the Hamiltonian
operator while $J^{\rho}_{(i)}$ are the components of the {\em
total} angular momentum operator. Moreover, the total angular
momentum is conserved in the sense that $[H,J_{(i)}^{\rho}]=0$.

Concluding we can say that, in our Cartesian gauge, the local
frames play the same role as the usual Cartesian {\rm rest frames}
of the central sources in flat spacetime since their axes are just
those of projections of the angular momenta.

\subsection{The diagonal gauge}

In other gauge fixings the basis-generators are quite different. A
tetrad gauge largely used in central charts with spherical
coordinates is the {\em diagonal}  gauge defined  by the 1-forms
\cite{DGB}
\begin{equation}\label{1fsf}
d\hat x^{0}_{s}=wdt\,,\quad
d\hat x^{1}_{s}=\frac{w}{u}dr\,,\quad
d\hat x^{2}_{s}=r\frac{w}{v}d\theta\,,\quad
d\hat x^{3}_{s}=r\frac{w}{v}\sin\theta d\phi\,.
\end{equation}
In this gauge the angular momentum operators of the canonical basis
(where $J_{(\pm)}=J_{(1)}\pm iJ_{(2)}$) are \cite{DGB}
\begin{equation}
J^{\rho}_{(\pm)}=L_{(\pm)}+\frac{e^{\pm i\phi}}{\sin\theta}\,\rho(S_{23})
\,,\quad
J^{\rho}_{(3)}=L_{(3)}\,.
\end{equation}
Thus one obtains a representation of $SU(2)$ where the spin terms
do not commute with the orbital ones and, therefore, the field
$\psi_{\rho}$ does not transform manifestly covariant under
rotations. In this case we can say that the spin part of the
central symmetry remains partially hidden  because of the diagonal
gauge which determines  special positions of the local frames with
respect to the natural one. However, when this is an impediment
one can change at anytime this gauge into the Cartesian one using
a simple local rotation. For the flat spacetimes these
transformations and their effects upon the Dirac equation are
studied in Ref. \cite{VIL}. We note that the form of the spin
generators as well as that of the mentioned rotation depend on the
enumeration of the 1-forms (\ref{1fsf}).

\section{The dS and AdS symmetries}

The backgrounds with highest external symmetry are the hyperbolic
manifolds, namely the dS and the AdS spacetimes. These are exact
solutions of the vacuum Einstein equations with cosmological
constant $\Lambda_c$. We shall briefly discuss simultaneously both
these manifolds which will be denoted by $M_{\epsilon}$ where
$\epsilon=1$ for dS case and $\epsilon=-1$ for AdS one. Our goal
here is to calculate the generators of the representations of the
group $S(M_{\epsilon})$ induced by those of $SL(2,\Bbb C)$.

The dS and AdS spacetimes are hyperboloids in the $(4+1)$ or
$(3+2)$-dimensional flat spacetimes, $M_{\epsilon}^{5}$, of
coordinates $Z^{A},\, A,B,...=0,1,2,3,5$, and the metric
$\eta(\epsilon)={\rm diag}(1,-1,-1,-1,-\epsilon)$. The equation of
the hyperboloid of radius $R=1/\hat\omega=\sqrt{3/|\Lambda_c|}$
reads
\begin{equation}\label{hip}
\eta_{AB}(\epsilon)Z^{A}Z^{B}=-\epsilon\,{R}^{2}\,.
\end{equation}
From their definitions it results that the dS or AdS spacetimes are
homogeneous spaces of the pseudo-orthogonal groups $SO(4,1)$ or
$SO(3,2)$ which play the role of gauge groups of the metric
$\eta(\epsilon)$ (for $\epsilon=1$ and  $\epsilon =-1$ respectively)
and represent just the isometry groups of these manifolds,
$G[\eta(\epsilon)]=I(M_{\epsilon})$. Then it is natural to use the
{\em covariant}  real parameters $\omega^{AB}=-\omega^{BA}$ since in
this parametrization the orbital basis-generators of the
representations of $G[\eta(\epsilon)]$, carried by the spaces of
functions over $M_{\epsilon}^{5}$, have the usual form
\begin{equation}\label{LAB5}
 ^{5}\!L_{AB}=i\left[\eta_{AC}(\epsilon)Z^{C}\partial_{B}-
 \eta_{BC}(\epsilon)Z^{C}
\partial_{A}\right].
\end{equation}
They  will give us directly the orbital basis-generators of the
representations of $S(M_{\epsilon})$ in the carrier spaces of the
functions defined over dS or AdS spacetimes. We note that in our
approach the generators of the groups $S(M_{\epsilon})$ are labeled
by same indices $(AB)$ as those of the groups $G[\eta(\epsilon)]$,
i.e. either $SO(4,1)$ or $SO(3,2)$.

\subsection{Central charts of AdS and dS spacetime}

The hyperboloid equation can be solved  in Cartesian dS/AdS coordinates,
$x^{0}=t$ and $x^{i}$ ($i=1,2,3$), which satisfy
\begin{eqnarray}\label{dScart}
Z^{5}\!\!&=&\hat\omega^{-1}\chi_{\epsilon}(r)\left\{\begin{array}{lll}
\cosh \hat\omega t&{\rm if}&\epsilon=1\\
\cos \hat\omega t&{\rm if}&\epsilon=-1
\end{array}\right.\nonumber\\
Z^{0}\!\!&=&\hat\omega^{-1}\chi_{\epsilon}(r)\left\{\begin{array}{lll}
\sinh \hat\omega t&{\rm if}&\epsilon=1\\
\sin \hat\omega t&{\rm if}&\epsilon=-1
\end{array}\right.\label{dsadsx}\\
Z^{i}&=& x^{i}\,,\nonumber
\end{eqnarray}
where we have denoted
$\chi_{\epsilon}(r)=\sqrt{1-\epsilon\,\hat\omega^{2}r^2}$. The
line elements
\begin{eqnarray}\label{dSmet}
ds^{2}&=&\eta_{AB}(\epsilon)dZ^{A}dZ^{B}\label{(adsm)}\\
&=&\chi_{\epsilon}(r)^{2} dt^{2}-
\frac{dr^{2}}{\chi_{\epsilon}(r)^{2}} -
 r^{2}(d\theta^{2}+\sin^{2}\theta\,d\phi^{2})\,,\nonumber
\end{eqnarray}
are defined on  the radial domains $D_{r}=[0,1/\hat\omega)$ or
$D_{r}=[0,\,\infty)$ for dS or AdS respectively.

We calculate the Killing vectors and the orbital generators of the
external symmetry in the Cartesian coordinates defined by Eq.
(\ref{dScart}) and the mentioned  parametrization of
$I(M_{\epsilon})$ starting with the identification
$\xi^{(AB)}=\omega^{AB}$. Then, from Eqs. (\ref{La}) and
(\ref{LAB5}), after a little calculation, we obtain the orbital
basis-generators
\begin{eqnarray}
L_{(05)}&=&\frac{i\epsilon}{\hat\omega}\,\partial_{t}\,,\\
L_{(j5)}&=&\frac{i\epsilon}{\hat\omega}\chi_{\epsilon}(r)\,
\left( \begin{array}{l}
{\rm cosh}\,\hat\omega t\\
\cos\hat\omega t
\end{array}\right)
\partial_{j}+\frac{ix^{j}}{\chi_{\epsilon}(r)}\,
\left(\begin{array}{l}
{\rm sinh}\,\hat\omega t\\
\sin\hat\omega t
\end{array}\right)
\partial_{t}\,,\\
L_{(0j)}&=&\frac{i}{\hat\omega}\chi_{\epsilon}(r)
\left(\begin{array}{l}
{\rm sinh}\,\hat\omega t\\
\sin\hat\omega t
\end{array}\right)
\partial_{j}+\frac{ix^{j}}{\chi_{\epsilon}(r)}
\left(\begin{array}{l}
{\rm cosh}\,\hat\omega t\\
\cos\hat\omega t
\end{array}\right)
\partial_{t}\,,\\
L_{(ij)}&=&-i\,(x^{i}\,\partial_{j}-x^{j}\,\partial_{i})\,.
\end{eqnarray}
Furthermore, we consider the Cartesian tetrad gauge defined by
Eqs. (\ref{(eee)})  - (\ref{(abc1)}) where, according to Eq.
(\ref{dSmet}), we have
\begin{equation}\label{dSuvw}
u(r)=\chi_{\epsilon}(r)^{2}\,,\quad
v(r)=w(r)=\chi_{\epsilon}(r)\,.
\end{equation}
 In this gauge we obtain the following local
$SL(2,\Bbb C)$ generators
\begin{eqnarray}
S_{(05)}(x)&=&0\,,\\
S_{(j5)}(x)&=&S_{0j}
\left(\begin{array}{l}
{\rm sinh}\,\hat\omega t\\
\sin\hat\omega t
\end{array}\right)
+\frac{1}{r^2}[\chi_{\epsilon}(r)
-1]\left[\epsilon\frac{S_{jk}x^{k}}{\hat\omega}
\left(\begin{array}{l}
{\rm cosh}\,\hat\omega t\\
\cos\hat\omega t
\end{array}\right)
\right.\nonumber\\
&&\left.
-\frac{S_{0k}x^{k}x^{j}}{\chi_{\epsilon}(r)}
\left(\begin{array}{l}
{\rm sinh}\,\hat\omega t\\
\sin\hat\omega t
\end{array}\right)
\right]\,,\\
S_{(0j)}(x)&=&S_{0j}
\left(\begin{array}{l}
{\rm cosh}\,\hat\omega t\\
\cos\hat\omega t
\end{array}\right)
+\frac{1}{r^2}[\chi_{\epsilon}(r)
-1]\left[\frac{S_{jk}x^{k}}{\hat\omega}
\left(\begin{array}{l}
{\rm sinh}\,\hat\omega t\\
\sin\hat\omega t
\end{array}\right)
\right.\nonumber\\
&&\left.
-\frac{S_{0k}x^{k}x^{j}}{\chi_{\epsilon}(r)}
\left(\begin{array}{l}
{\rm cosh}\,\hat\omega t\\
\cos\hat\omega t
\end{array}\right)
\right]\,,\\
S_{(ij)}(x)&=&S_{ij}\,.
\end{eqnarray}
With their help we can write the action of the spin terms
(\ref{sss}) and, implicitly, that of the  basis-generators
$X_{(AB)}^{\rho}=L_{(AB)}+S_{(AB)}^{\rho}$ of the representations of
$S(M_{\epsilon})$ induced by the representations  $\rho$ of
$SL(2,\Bbb C)$. Hereby it is not difficult to show that
$S(M_{\epsilon})$ is isomorphic with the universal covering group of
$I(M_{\epsilon})$ which in both cases ($\epsilon=\pm1$) is a
subgroup of the $SU(2,2)$ group.

As  expected, in central charts and Cartesian gauge the fields
transform manifestly covariant only under the transformations of the
subgroup $SU(2)\subset S(M_{\epsilon})$. Consequently, the total
angular momentum has the usual form given by Eq. (\ref{ang1}), with
point independent spin terms. The hamiltonian operator $H\propto
L_{(05)}$ has no spin parts but all the other generators have spin
parts depending on coordinates.

\subsection{Minkowskian charts}

Another possibility is to solve the hyperboloid equation (\ref{hip})
in Minkowskian charts \cite{SW} where the coordinates, $x^{\mu}$,
are defined by
\begin{equation}
Z^{5}={\hat\omega}^{-1}\tilde\chi_{\epsilon}(s)\quad Z^{\mu}=x^{\mu}
\end{equation}
with
$\tilde\chi_{\epsilon}(s)=\sqrt{1+\epsilon\,\hat\omega^{2}s^{2}}$
and $s^{2}=\eta_{\mu\nu}x^{\mu}x^{\nu}$. In these coordinates it is
convenient to identify the hat indices with the usual ones and to do
not raise or lower these indices. Then we find that the metric
tensor,
\begin{equation}
g_{\mu\nu}(x)=\eta_{\mu\nu}-\frac{\epsilon\,\hat\omega^2}
{\tilde\chi_{\epsilon}(s)^2}
\eta_{\mu\alpha}x^{\alpha}\eta_{\nu\beta}x^{\beta}
\end{equation}
transforms manifestly covariant under the global $L_{+}^{\uparrow}$
transformations, $x'^{\mu}\to
x^{\mu}=\Lambda^{\mu\,\cdot}_{\cdot\,\nu}x^{\nu}$. Moreover, the
whole theory remains manifest covariant if we use the tetrad fields
in the Lorentz gauge defined as
\begin{equation}
e^{\mu}_{\nu}(x)=\delta^{\mu}_{\nu}+h_{\epsilon}(s)\eta_{\nu\alpha}x^{\alpha}
x^{\mu}\quad \hat e^{\mu}_{\nu}(x)=\delta^{\mu}_{\nu}+\hat
h_{\epsilon}(s)\eta_{\nu\alpha} x^{\alpha}x^{\mu}
\end{equation}
where
\begin{equation}
h_{\epsilon}(s)=\frac{1}{s^2}\left[\tilde\chi_{\epsilon}(s)-1
\right]\quad \hat
h_{\epsilon}(s)=\frac{1}{s^2}\left[\frac{1}{\tilde\chi_{\epsilon}(s)}-1
\right]\,.
\end{equation}

First we calculate the  $SO(4,1)$ or $SO(3,2)$ orbital generators,
\begin{eqnarray}
L_{(\mu5)}&=&\frac{i\epsilon}{\hat\omega}\tilde\chi_{\epsilon}(s)
\partial_{\mu}\\
L_{(\mu\nu)}&=&i(\eta_{\mu\alpha}x^{\alpha}\partial_{\nu}
-\eta_{\nu\alpha}x^{\alpha}\partial_{\mu})
\end{eqnarray}
which are independent on the gauge fixing. We observe that in
Minkowskian charts $\partial_{t}$ is no more a Killing vector field
as in the case of the central ones. However, here we have another
advantage, namely that of the Lorentz gauge in which the local
$sl(2,\Bbb C)$ generators of Eq.(\ref{sss}) have the form
\begin{eqnarray}
S_{(\mu5)}(x)&=&-\frac{\epsilon}{\hat\omega s^2}\left[
\tilde\chi_{\epsilon}(s)-1\right]
S_{\mu\alpha}x^{\alpha}\\
S_{(\mu\nu)}(x)&=&S_{\mu\nu}
\end{eqnarray}
showing that the field $\psi_{\rho}$ transforms manifestly covariant
under the whole $SL(2,\Bbb C)$ subgroup of $S(M_{\epsilon})$. Since
these representations are induced just by those of $SL(2,\Bbb C)$ we
can say that in this gauge the manifest covariance is maximal.

\section{The Dirac field in curved backgrounds}

In what follows we study the Dirac particles {\em freely} moving
on dS or AdS backgrounds without to affect the geometry. The
problems of this type are  called often Dirac perturbations on a
given manifold.

The main purpose is to find the quantum modes of the Dirac field
determined by complete sets of commuting operators. The Noether
theorem applied to our theory of external symmetry gives us
classical conserved quantities or conserved operators in quantum
theory, corresponding to the generators of the group $S(M)$. Thus,
the quantum theory is equipped with a large algebra of conserved
operators among them we can select the systems of commuting
operators we need for defining the quantum modes which will help us
to perform the second quantization of the Dirac field in canonical
manner.

\subsection{The Dirac field}

The form of the Dirac field in curves backgrounds depends on two
basic elements: the choice of the tetrad gauge and the
representation of the Dirac's $\gamma$-matrices. These satisfy
\begin{equation}
\{ \gamma^{\hat\alpha}, \gamma^{\hat\beta} \}=2\eta^{\hat\alpha
\hat\beta}
\end{equation}
and give the generators of the reducible spinor representation of
the $SL(2,C)$ group \cite{TH} we denote from now directly by
$S^{\hat\alpha \hat\beta}$ instead of $\rho(S^{\hat\alpha
\hat\beta})$. These generators,
\begin{equation}
S^{\hat\alpha \hat\beta}=\frac{i}{4} [\gamma^{\hat\alpha},
\gamma^{\hat\beta} ]\,,
\end{equation}
are self-adjoint, $\overline{S}^{\hat\alpha\hat\beta}
=S^{\hat\alpha \hat\beta}$, (with respect to the Dirac adjoint
defined as $\overline X=\gamma^0 X^+\gamma^0$) and satisfy
\begin{eqnarray}
[S^{\hat\alpha\hat\beta},\,\gamma^{\hat\sigma}]&=&
i(\eta^{\hat\beta\hat\sigma}\gamma^{\hat\alpha}-
\eta^{\hat\alpha\hat\sigma}\gamma^{\hat\beta})\,,\label{Sgg}\\
{[} S_{\hat\alpha\hat\beta},\,S_{\hat\sigma\hat\tau} {]}&=&i(
\eta_{\hat\alpha\hat\tau}\,S_{\hat\beta\hat\sigma}-
\eta_{\hat\alpha\hat\sigma}\,S_{\hat\beta\hat\tau}+
\eta_{\hat\beta\hat\sigma}\,S_{\hat\alpha\hat\tau}-
\eta_{\hat\beta\hat\tau}\,S_{\hat\alpha\hat\sigma})\,,\label{SSS}
\end{eqnarray}
In the chiral representation of the spinor field the crucial role is
played by the matrix $\gamma^5=-i\gamma^0\gamma^1\gamma^2\gamma^3$
which is diagonal in this representation. We note that the set of
matrices $\{
S^{\hat\alpha\hat\beta},\gamma^{\hat\mu},\gamma^{\hat\mu}\gamma^5,\gamma^5\}$
form the basis of a fundamental representation of the $su(2,2)$
algebra \footnote{For extended bibliography see Ref. \cite{SU22}}.

The theory of the Dirac field $\psi$ of mass $m$,  defined on the
space domain $D$, is based on the gauge invariant action \cite{BD},
\begin{equation}\label{action}
{\cal S}[\psi]=\int\, d^{4}x\sqrt{g}\left\{
\frac{i}{2}[\bar{\psi}\gamma^{\hat\alpha}D_{\hat\alpha}\psi-
(\overline{D_{\hat\alpha}\psi})\gamma^{\hat\alpha}\psi] -
m\bar{\psi}\psi\right\}
\end{equation}
where $\overline \psi=\psi^+\gamma^0$ is the Dirac adjoint of
$\psi$ and
\begin{equation}
D_{\hat\alpha}=\hat\partial_{\hat\alpha}+\frac{i}{2}S^{\hat\beta
\cdot}_{\cdot \hat\gamma}\hat\Gamma^{\hat\gamma}_{\hat\alpha
\hat\beta}
\end{equation}
are the covariant derivatives of the spinor field given by Eq.
(\ref{covder}). The Dirac equation resulted from the action
(\ref{action}) reads
\begin{equation}\label{EEE}
 E_D\psi-m\psi=0\,, \quad
 E_D=i\gamma^{\hat\alpha}D_{\hat\alpha}\,.
\end{equation}
The Dirac operator $E_D$ can be put in a more comprehensive form as
\begin{equation}\label{(dd)}
E_D=i\gamma^{\hat\alpha}e_{\hat\alpha}^{\mu}\partial_{\mu}  +
\frac{i}{2}
\frac{1}{\sqrt{g}}\partial_{\mu}(\sqrt{g}e_{\hat\alpha}^{\mu})
\gamma^{\hat\alpha} -\frac{1}{4} \{\gamma^{\hat\alpha},
S^{\hat\beta \cdot}_{\cdot \hat\gamma} \}
\hat\Gamma^{\hat\gamma}_{\hat\alpha \hat\beta}\,.
\end{equation}
In some static problems where $\hat e^0_i=0,\, i=1,2,3$, it is
convenient to put the Dirac equation in Hamiltonian form,
$i\partial_t\psi=H_D\psi$, where the Hamiltonian operator, $H_D$,
can be calculated from Eqs. (\ref{EEE}) and (\ref{(dd)}).

On the other hand, from the conservation of the electric charge, one
can define the time-independent relativistic scalar product of two
spinors \cite{BD}. When $e^{0}_{i}=0$, $i=1,2,3$, this takes the
simple form
\begin{equation}\label{sp}
\left<\psi,\psi'\right>=\int_{D}d^{3}x\,\mu(x)\bar\psi(x)\gamma^{0}\psi'(x)\,,
\quad
\end{equation}
where
\begin{equation}\label{(weight)}
\mu(x)=\sqrt{g(x)}\,e_{0}^{0}(x)
\end{equation}
is the specific weight function of the Dirac field. In the central
chart $\{t,\vec{x}\}$ with the metric (\ref{(metr)}) this weight
function reads
\begin{equation}\label{(mu)}
\mu(\vec{x}) =\frac{1}{b^{2}(b+r^{2}c)} =\frac{w^3}{uv^2}
\end{equation}
since $g$ is given by Eq. (\ref{ggg}).

Our theory of external symmetry offers us the framework we need to
calculate the conserved quantities predicted by the Noether
theorem. Starting with the infinitesimal transformations of the
one-parameter subgroup of $S(M)$ generated by $X_a$ corresponding
to the Killing vector $k_a$, we find that there exists the
conserved current $\Theta^{\mu}[X_a]$ which satisfies
$\Theta^{\mu}[X_a]_{;\mu}=0$. Using the  action (\ref{action}) we
obtain the concrete form of these currents,
\begin{equation}
\Theta^{\mu}[X_a]=-\tilde T^{\mu\,\cdot}_{\cdot\,\nu}k_a^{\nu}+
\frac{1}{4} \,\overline{\psi}\{\gamma^{\hat\alpha}, S^{\hat\beta
\hat\gamma} \}\psi\, e^{\mu}_{\hat\alpha} \,\Omega_{a\,\hat\beta
\hat\gamma}\,,
\end{equation}
where the functions $\Omega_{a\,\hat\beta \hat\gamma}$ are defined
by Eq. (\ref{Om}) while
\begin{equation}
\tilde T^{\mu\,\cdot}_{\cdot\,\nu}=
\frac{i}{2}\left[\overline{\psi}\gamma^{\hat\alpha}e^{\mu}_{\hat\alpha}
\partial_{\nu}\psi-
(\overline{\partial_{\nu}\psi})\gamma^{\hat\alpha}e^{\mu}_{\hat\alpha}\psi
\right]
\end{equation}
is a notation for a part of the stress-energy tensor of the Dirac
field \cite{SW,BD}. Finally, it is clear that the conserved
quantity corresponding to the Killing vector $k_a$ is the real
number
\begin{equation}\label{cq}
\int_{D} d^3 x \sqrt{g}\,\Theta^{0}[X_a]=
\frac{1}{2}\left[\left<\psi, X_a\psi\right>+\left<X_a\psi,
\psi\right>\right]\,.
\end{equation}
We note that it is premature to interpret this formula  as an
expectation value or to speak about Hermitian conjugation of the
operators $X_a$ with respect to the scalar product (\ref{sp}),
before specifying the boundary conditions on $D$.

What is important here is that this result is useful in
quantization giving directly the one-particle operators of the
quantum field theory when $\psi$ becomes a field operator. Indeed,
starting with the form (\ref{cq}) of the conserved quantities, we
find that for any self-adjoint generator $X$ of the spinor
representation of the group $S(M)$ there exists a {\em conserved}
one-particle operator of the quantum field theory which can be
calculated simply as
\begin{equation}\label{opo}
{\bf X}=:\left<\psi, X\psi\right>:
\end{equation}
respecting the normal ordering  ($::$) of the operator products
\cite{BJD}. The quantization rules must be postulated such that the
standard algebraic properties
\begin{equation}\label{algXX}
[{\bf X}, \psi(x)]=-X\psi(x)\,, \quad [{\bf X}, {\bf
X}']=:\left<\psi, [X,X']\psi\right>:
\end{equation}
should be accomplished.

\subsection{The reduced Dirac equation in central charts}

In what follows we consider  manifolds $M$ with central symmetry
having either charts  with Cartesian coordinates $\{t,\vec{x}\}$
and the line element (\ref{(metr)}) or charts with spherical
coordinates $\{t,r,\theta,\phi\}$ where we prefer to work with
metrics of the form (\ref{(muvw)}). In these charts we chose the
Cartesian gauge defined by Eqs. (\ref{(eee)}), (\ref{(eee1)}),
(\ref{(abc)}) and (\ref{(abc1)}) which bring the rotations in
manifest covariant form.

The Dirac equation in central charts can be written replacing the
concrete form of the tetrad components in Eq. (\ref{(dd)}). First,
we observe that the last term of this equation does not contribute
when the metric is spherically symmetric. The argument is that
$\{\gamma^{\hat\alpha},S^{\hat\beta \hat\gamma}\}=
\varepsilon^{\hat\alpha \hat\beta \hat\gamma \cdot}_{~~~
\hat\lambda} \gamma^{5}\gamma^{\hat\lambda}$ (with
$\varepsilon^{0123}=1$) is completely antisymmetric, while the
Cartan coefficients resulted from Eqs. (\ref{(eee)}) and
(\ref{(eee1)}) have no such type of components. Furthermore,  we
bring the remaining equation in a simpler form defining the {\it
reduced} Dirac field, $\tilde\psi$, as
\begin{equation}\label{(cfu)}
\psi(x)=\chi(r)\tilde\psi(x).
\end{equation}
where
\begin{equation}\label{(chi)}
\chi=[\sqrt{g}(b+r^{2}c)]^{-1/2}=b\sqrt{a}=vw^{-3/2}\,.
\end{equation}
In this way  we obtain the {\em reduced} Dirac equation in the
central chart $(t,\vec{x})$ with Cartesian coordinates and
Cartesian tetrad gauge,
\begin{equation}\label{(red)}
i\left\{a(r)\gamma^{0}\partial_{t}
+b(r)(\stackrel{\rightarrow}{\gamma}\cdot
\stackrel{\rightarrow}{\partial})+
c(r)(\stackrel{\rightarrow}{\gamma}\cdot
\stackrel{\rightarrow}{x})[1+ (\stackrel{\rightarrow}{x}\cdot
\stackrel{\rightarrow}{\partial})]\right\}\tilde\psi(x)
=m\tilde\psi(x)\,.
\end{equation}
We observe that this equation is invariant under time translations
which means that the energy is conserved. For this reason it is
convenient to bring it in Hamiltonian form,
\begin{equation}\label{HHD}
{\tilde H}_D\tilde\psi = i\partial_t\tilde\psi\,,
\end{equation}
using the operators
\begin{eqnarray}
{\tilde H_D}&=&-i\frac{u(r)}{r^2}(\gamma^0\gamma^i x^i)\left( 1+
x^i\partial_i\right)-i\frac{v(r)}{r^2}(\gamma^i x^i){K}
+w(r)\gamma^0 m\,,\label{HHH}\\
{K}&=&\gamma^0 \left(2\vec{S}\cdot \vec{L} +1\right)\,.\label{KKK}
\end{eqnarray}
We note that the reduced Hamiltonian operator $\tilde
H_D=\chi^{-1}H_D\chi$ is related to the original Hamiltonian $H_D$
corresponding to $E_D$.

The operator $K$ which concentrates all the angular terms of $\tilde
H_D$ as well as the total angular momentum $\vec{J}=\vec{L}+\vec{S}$
defined by Eq. (\ref{ang1}) commute with $\tilde H_D$ and satisfy
$K^2=\vec{J}^2+\frac{1}{4}$. Consequently, all the properties
related to the conservation of the angular momentum, including the
separation of variables in spherical coordinates, will be similar to
those of the usual Dirac theory in Minkowski spacetimes. Thus in Eq.
(\ref{HHD}) one can separate the spherical variables with the help
of the four-components angular spinors $\Phi^{\pm}_{m_{j},
\kappa_{j}}$ used in the central problems  of special relativity
\cite{TH}. These are eigenspinors of the complete set
$\{\vec{J}^2,K,J_3 \}$ corresponding to the eigenvalues $\{j(j+1),
-\kappa_j,m_j\}$ where $\kappa_{j}$ can take only the values $\pm
(j+1/2)$. We note that for each set of quantum numbers $(j,\kappa_j,
m_j)$ there are two orthogonal spinors, denoted by the superscripts
$\pm$, as defined in Ref. \cite{TH}.

We must specify that in curved manifolds with central symmetry this
conjecture can be obtained only in our Cartesian tetrad gauge where
the representations of the group $SU(2)$ are manifest covariant
since $[L_i,S_j]=0$. In other gauge fixings where $L_i$ and $S_i(x)$
do not commute among themselves the form of Eqs. (\ref{HHH}) and
(\ref{KKK}) may be different \cite{DGB,SOL2} and then the spinors
$\Phi^{\pm}_{m_{j}, \kappa_{j}}$ become useless. For example, in the
diagonal gauge defined by Eq. (\ref{1fsf}) one has to consider
another type of angular spinors, constructed with the help of the so
called spin-weighted harmonics \cite{PN}.

In other respects, we can verify that in our gauge the discrete
transformations, $P$, $C$ and $T$, have the same significance and
action as those of special relativity \cite{BJD,TH}. This is because
the form of the Hamiltonian operator (\ref{HHH}) is close to that
meet in the flat case. We shall see later that the charge
conjugation transforms each particular solution of positive
frequency of Eq. (\ref{(red)}) into the corresponding one of
negative frequency.

\subsection{The radial problem}

Our goal is to write down the eigenspinors of the complete set of
commuting operators $\{ H_D, \vec{J}^2,K,J_3\}$ corresponding to the
eigenvalues $\{E,j(j+1),-\kappa_j,m_j\}$. These are particular
solutions of the original Dirac equation (\ref{EEE}) in the central
chart $\{t,r,\theta,\phi\}$ with the line element (\ref{(muvw)}). We
consider first the solutions of  positive frequency and the energy
$E$ that read,
\begin{eqnarray}
&&U_{E,j,\kappa_{j},m_{j}}({x})=U_{E,j,\kappa_{j},m_{j}}(t,r,\theta,\phi)\nonumber\\
&&~~=\frac{v(r)}{r w
(r)^{3/2}}[f^{(+)}_{E,\kappa_j}(r)\Phi^{+}_{m_{j},\kappa_{j}}(\theta,\phi)
+f^{(-)}_{E,\kappa_j}(r)\Phi^{-}_{m_{j},\kappa_{j}}(\theta,\phi)]e^{-iEt}\,.\label{(psol)}
\end{eqnarray}
The reduced part of these solutions, $\tilde U=U/\chi$, satisfy
Eq. (\ref{HHD}) from which, after the separation of angular
variables, we obtain the radial equations
\begin{eqnarray}
\left[u(r)\frac{d}{dr}+v(r)\frac{\kappa_j}{r}\right]f^{(+)}(r)&=&[E+w(r)m
]f^{(-)}(r),\label{(e1)}\\
\left[-
u(r)\frac{d}{dr}+v(r)\frac{\kappa_j}{r}\right]f^{(-)}(r)&=&[E-w(r)m
]f^{(+)}(r)\,,\label{(e2)}
\end{eqnarray}
where we omitted the indices $E$ and $\kappa_j$ of the radial
functions. In practice, these radial equations can be written
directly starting with the line element put in the form Eq.
(\ref{(muvw)}) from which we can identify  the functions $u$, $v$,
and $w$ we need.

The normalization of the Dirac spinors has to be done with respect
to the scalar product (\ref{sp}). In the chart $\{t,r,\theta,\phi\}$
the weight function (\ref{(weight)}) becomes
\begin{equation}\label{(mus)}
\mu(r,\theta)=\frac{w^3}{uv^2}r^2\sin(\theta)
\end{equation}
and, consequently, the integral (\ref{sp}) splits in radial and
angular terms. The angular spinors $\Phi^{\pm}_{m_{j},\kappa_{j}}$
are orthogonal and normalized to unity with respect to their own
angular scalar product defined on the sphere $S^2$ by the angular
integral \cite{TH}. In this way we are left with the radial integral
\begin{eqnarray}
\left<U_1,U_2\right>&=&\delta_{\kappa_{j_1},\kappa_{j_2}}\delta_{m_{j_1},m_{j_2}}\nonumber\\
&&\times \int_{D_{r}}\frac{dr}{u(r)}\{[f_{1}^{(+)}(r)]^{*}
f_{2}^{(+)}(r)+[f_{1}^{(-)}(r)]^{*}f_{2}^{(-)}(r)\}\,, \label{(spp)}
\end{eqnarray}
calculated according to  Eqs. (\ref{(cfu)}) and (\ref{(psol)}). The
domain $D_{r}$ is the radial domain corresponding to $D$. What is
remarkable here is that the radial weight function $1/u$, resulted
from Eqs. (\ref{(mu)}) and (\ref{(chi)}), guarantees that
$(u\partial_{r})^{+}=-u\partial_{r}$. This means that the operators
of the left-hand side of the radial equations are related between
themselves through the Hermitian conjugation with respect to the
scalar product (\ref{(spp)}).

The principal consequence is that we can define the self-adjoint
{\em radial} Hamiltonian
\begin{equation}
H_r=\begin{array}{|cc|}
    mw& -u\frac{\textstyle d}{\textstyle dr}+\kappa_{j}\frac{\textstyle v}
{\textstyle r}\\
       u\frac{\textstyle d}{\textstyle dr}+\kappa_{j}\frac{\textstyle v}
{\textstyle r}& -mw
\end{array}\,.
\end{equation}
This operator allows us to write the Eqs. (\ref{(e1)}) and
(\ref{(e2)}) as the eigenvalue problem
\begin{equation}\label{(hfef)}
H_r{\cal F}=E{\cal F},
\end{equation}
where the two-dimensional eigenvectors  ${\cal F}=(f^{(+)},
f^{(-)})^{T}$ have their own radial scalar product,
\begin{equation}\label{(spf)}
({\cal F}_{1},{\cal F}_{2})=\int_{D_{r}}\frac{dr}{u(r)} {\cal
F}_{1}^{+}{\cal F}_{2}\,,
\end{equation}
as we deduce from Eq. (\ref{(spp)}). Thus we obtained an {\em
independent} radial problem which has to be solved in  each
particular case separately using appropriate methods.

First of all, we  look for  possible transformations which should
simplify the radial equations. It is known that the
transformations of the space coordinates of a natural frame with
static metric do not change the quantum modes.  In our case we can
change only the radial coordinate without to affect the central
symmetry. A good choice is the chart where the radial coordinate
is defined by
\begin{equation}\label{(spfr)}
r_{s}(r)=\int\frac{dr}{u(r)}+const
\end{equation}
so that  $r_{s}(0)=0$.  This chart will be called  {\it special}
central chart (frame). In the following we shall use only this
frame starting with metrics with $u=1$ while the subscript $s$
will be omitted.

In radial problems of the Dirac theory in flat spacetime
\cite{TH} the manifest supersymmetry play an important role in
solving the radial equations. For this reason we look for
transformations of the form
 ${\cal F}\to \hat {\cal F}=U{\cal F}$ and $H_r\to \hat H_r=UH_rU^{-1}$ which
 could simplify the radial problem pointing out a {\em hidden} supersymmetry.
When this exists, the transformed Hamiltonian must take the standard
supersymmetric form
\begin{equation}\label{(ssh)}
\hat H_r=\begin{array}{|cc|}
    \nu& -\frac{\textstyle d}{\textstyle dr}+W\\
       \frac{\textstyle d}{\textstyle dr}+W& -\nu
\end{array}
\end{equation}
where $\nu$ is a constant and $W$ is the resulting superpotential
\cite{COTA}. If the radial problem has this property, then the
second order equations for the components $\hat f^{(+)}$ and $\hat
f^{(-)}$ of $\hat{\cal F}$ can be obtained from $\hat
{H_r}^{2}\hat{\cal F}=E^{2}\hat{\cal F}$. These equations,
\begin{equation}\label{(fpm)}
\left(-\frac{d^{2}}{dr^2}+W(r)^{2}\mp
\frac{dW(r)}{dr}+\nu^{2}\right) \hat f^{(\pm)}(r)=E^{2}\hat
f^{(\pm)}(r),
\end{equation}
represent the starting point for finding analytical solutions.

The simplest radial problems are those with manifest
supersymmetry, for which the original radial Hamiltonian $H_r$ has
the form (\ref{(ssh)}). These are generated by the metrics of the
central manifolds ${\Bbb R}\times M_{3}$ which have $w=1$.  In the
special frames (where $u=1$) these  are  determined only by the
arbitrary function $v$  which gives the superpotential
$W=\kappa_{j}v/r$. The most popular example is the
three-dimensional sphere with the time trivially added, ${\Bbb
R}\times S^3$. A more complicated situation is when  we need to
use a suitable transformation $U$ in order to point out the
supersymmetry.  These are problems with  hidden supersymmetry
which are similar with that of the Dirac particle in external
Coulomb field, known from special relativity. However,  examples
of radial problems  in which the supersymmetry is much more hidden
will be discussed in the next section.

\section{Dirac field in cental charts of the AdS and dS spacetimes}

In our approach  the Dirac equation in the central charts (frames)
of AdS or dS backgrounds can be analytically solved in terms of
Gauss hypergeometric functions \cite{COTA,COTA1}. One obtains thus
the fundamental solutions giving the quantum modes. Since the
central frames are static, these solutions are energy eigenspinors
corresponding to discrete or continuous energy spectra. When these
spinors can be normalized (in usual or generalized sense) then the
canonical quantization of the Dirac field in these frames may be
done.

\subsection{The Dirac quantum field in AdS spacetime}

We consider first the AdS spacetime and we present how can be
selected the quantum modes of the Dirac field and to write down the
normalized energy eigenspinors of the regular modes. The final
objective is to quantize this field \cite{COTA}.

The AdS spacetime of radius $R$ is defined by Eq. (\ref{hip}) for
$\epsilon=-1$. Changing the notation, we denote from now
$\omega=1/R$ instead of $\hat\omega$. This manifold has a special
central chart $\{t,r,\theta,\phi\}$ with the line element \cite{AIS}
\begin{equation}\label{(le)}
ds^{2}=\sec^{2}\omega r \left[dt^{2}-dr^{2}-\frac{1}{\omega^{2}}
\sin^{2}\omega r~ (d\theta^{2}+\sin^{2}\theta~d\phi^{2})\right]\,.
\end{equation}
The radial domain of this chart is $D_{r}=[0,\pi/2\omega)$ because
of the event horizon at $r=\pi/2\omega$. We specify that here we
take $t\in (-\infty, \infty)$ which defines in fact the universal
covering spacetime (CAdS) of AdS \cite{AIS}.

Now, from Eq. (\ref{(le)}) we can identify the functions
$w(r)=\sec \omega r$ and $v(r)=\omega r \csc \omega r$. With their
help and using the notation $k=m/\omega$  we obtain the radial
Hamiltonian,
\begin{equation}
H_r=\begin{array}{|cc|}
    \omega k\sec \omega r& -\frac{\textstyle d}{\textstyle dr}+
\omega\kappa_{j}\csc\omega r\\
&\\
       \frac{\textstyle d}{\textstyle dr}+\omega\kappa_{j}\csc\omega r
& -\omega k\sec \omega r
\end{array}\,,
\end{equation}
which has to give us the radial functions of the particular
solutions (\ref{(psol)}).

Despite of the fact that this is not obvious at all, this
Hamiltonian has a hidden supersymmetry that can be pointed out
with the help of the local rotation ${\cal F}\to \hat{\cal
F}=R{\cal F}=(\hat f^{(+)},\hat f^{(-)})^{T}$ produced by
\begin{equation}\label{(uder)}
R(r)=\begin{array}{|cc|}
    \cos \frac{\textstyle \omega r}{\textstyle 2}&-\sin
\frac{\textstyle \omega r}{\textstyle 2}\\
&\\
\sin \frac{\textstyle \omega r}{\textstyle 2}&\cos
\frac{\textstyle \omega r}{\textstyle 2}
\end{array}\,.
\end{equation}
Indeed, after a few manipulations we find that the transformed
(i.e. rotated and translated) Hamiltonian,
\begin{equation}\label{(newh)}
\hat H_r =RH_rR^{T}-\frac{\omega}{2} 1_{2\times 2},
\end{equation}
has supersymmetry since it  has  the requested specific form
(\ref{(ssh)}) with diagonal constant terms $\nu=\omega(k-\kappa_j)$
and the superpotential \cite{COTA},
\begin{equation}
W(r)=\omega(k\tan(\omega r)+\kappa_j\cot(\omega r))\,,
\end{equation}
which is of the P\" oschl-Teller type \cite{PT} .

In these circumstances the new eigenvalue problem
\begin{equation}\label{(trrp)}
\hat H_r \hat{\cal F}=\left(E-\frac{\omega}{2}\right)\hat{\cal F},
\end{equation}
involving the transformed radial wave functions $\hat f^{(\pm)}$,
leads to a pair of  second order equations
\begin{equation}
\left(-\frac{d^2}{dr^2}+\omega^{2}\frac{k(k\mp 1)}{\cos^{2}\omega
r}+ \omega^{2}\frac{\kappa_{j}(\kappa_{j}\pm 1)}{\sin^{2}\omega
r}\right) \hat f^{(\pm)}(r)=
\omega^{2}\epsilon^{2}\hat f^{(\pm)}(r)\label{(od1)},\\
\end{equation}
where we denoted $\epsilon=E/\omega-1/2$. These equations have
analytical solutions \cite{COTA},
\begin{eqnarray}
\hat f^{(\pm)}(r)&=&N_{\pm}
\sin^{2s_{\pm}}\omega r\cos^{2p_{\pm}}\omega r \label{(gsol)}\\
&&\times F\left(s_{\pm}+p_{\pm}-\frac{\epsilon}{2},
s_{\pm}+p_{\pm}+\frac{\epsilon}{2}, 2s_{\pm}+\frac{1}{2},
\sin^{2}\omega r \right).\nonumber
\end{eqnarray}
where  $F$ are Gauss hypergeometric functions \cite{AS}. Their real
parameters are defined as
\begin{eqnarray}
2s_{\pm}(2s_{\pm}-1)&=&\kappa_{j}(\kappa_{j}\pm 1)\,,\label{(2s)}\\
2p_{\pm}(2p_{\pm}-1)&=&k(k\mp 1)\,,\label{(2p)}
\end{eqnarray}
while $N_{\pm}$ are normalization factors. The next step is to
select the suitable values of these parameters  and to calculate
$N_{+}/N_{-}$ such that the functions $\hat f^{(\pm)}$ should be
solutions of the transformed radial problem (\ref{(trrp)}), with a
good physical meaning. This can be achieved only when $F$ is a
polynomial selected by a suitable quantization condition since
otherwise $F$ is strongly divergent for $\sin^{2}\omega r\to 1$.
Then the  functions $\hat f^{(\pm)}$ will be square integrable and
the normalization factors can be calculated according to the
condition
\begin{equation}\label{(norm)}
({\cal F},{\cal F})=
 (\hat{\cal F},\hat{\cal F})
=\int_{D_{r}}dr\,\left( |\hat f^{(+)}(r)|^{2}+ |\hat
f^{(-)}(r)|^{2}\right)=1\,,
\end{equation}
resulted from the fact that the matrix (\ref{(uder)}) is
orthogonal.

The discrete energy spectrum is given by the  particle-like CAdS
quantization conditions
\begin{equation}\label{(quant)}
\epsilon=2 (n_{\pm}+s_{\pm}+p_{\pm})\,, \quad \epsilon>0\,,
\end{equation}
that must be compatible with each other, i.e.
\begin{equation}\label{(comp)}
n_{+}+s_{+}+p_{+}=n_{-}+s_{-}+p_{-}.
\end{equation}
Hereby we see  that there is only one independent {\em radial}
quantum number, $n_{r}=0,1,2,...$. In addition, we use  the orbital
quantum number $l$ of the spinor $\Phi^{+}_{m_{j},\kappa_{j}}$
\cite{TH}, as an auxiliary quantum number. On the other hand, if we
express  (\ref{(gsol)}) in terms of Jacobi polynomials, we observe
that these functions remain square integrable for $2s_{\pm}>-1/2$
and $2p_{\pm}>-1/2$. Since $l=0,1,2...$ we are forced to select only
the positive solutions of Eqs. (\ref{(2s)}). The different solutions
of Eqs. (\ref{(2p)}) defines the boundary conditions of the allowed
quantum modes, like in the case of scalar modes \cite{SOL1}. We say
that for $k>-1/2$ the values $2p_{+}=k$ and $2p_{-}=k+1$ define the
boundary conditions of {\em regular} modes. The other possible
values, $2p_{+}=-k+1$ and $2p_{-}=-k$, define the {\em irregular}
modes when $k<1/2$. Obviously, for $-1/2<k<1/2$ both these modes are
possible.

We note that the AdS quantization conditions require, in addition,
$k$ to be a half integer. Then it is clear that the domains of $k$
corresponding to the regular and respectively irregular modes can
not overlap with each other. Anyway, in our opinion, the problem of
the meaning of the irregular modes as well as that of the relation
between these kind of modes is sensitive and may be carefully
analyzed. For this reason we restrict ourselves to write down only
the energy eigenspinors of the regular modes on CAdS.

Let us  take first $\kappa_{j}=-(j+1/2)=-l-1$. Then the positive
solutions of (\ref{(2s)}) are $2s_{+}=l+1$ and $2s_{-}=l+2$ while,
according to (\ref{(comp)}), we must have $n_{+}=n_{r}$ and
$n_{-}=n_{r}-1$. For these values of  parameters, the functions
$\hat f^{(\pm)}$ given by Eqs. (\ref{(gsol)}) and (\ref{(quant)})
represent a correct solution of the transformed radial problem
(\ref{(trrp)}) only if
\begin{equation}\label{(npnm)}
\frac{N_{-}}{N_{+}}=-\frac{2n_{r}}{2l+3}.
\end{equation}
Furthermore, it is easy to express (\ref{(gsol)}) in terms of
Jacobi polynomials and to calculate the normalization factors
according to Eq. (\ref{(norm)}). Thus we arrive at the result,
\begin{eqnarray}
\hat f^{(+)}(r)_{|\kappa_{j}=-(j+1/2)}&=&
N\left[\frac{n_{r}+k+l+1}{n_{r}+l+\frac{1}{2}}\right]^{\frac{1}{2}}\nonumber\\
&&\times \sin^{l+1}\omega r \cos^{k}\omega r
P_{n_{r}}^{(l+\frac{1}{2},k-\frac{1}{2})}(\cos 2\omega r)\,,
\label{(1)}\\
\hat f^{(-)}(r)_{|\kappa_{j}=-(j+1/2)}&=&-N
\left[\frac{n_{r}+k+l+1}{n_{r}+l+
\frac{1}{2}}\right]^{\frac{1}{2}}\nonumber\\
&&\times \sin^{l+2}\omega r \cos^{k+1}\omega r
P_{n_{r}-1}^{(l+\frac{3}{2},k+\frac{1}{2})}(\cos 2\omega r)\,,
\nonumber
\end{eqnarray}
where
\begin{equation}
N=\eta\sqrt{2\omega}\left[\frac{n_{r}!\,\Gamma(n_{r}+k+l+1)}
{\Gamma(n_{r}+l+\frac{1}{2})\Gamma(n_{r}+k+\frac{1}{2})}\right]^{\frac{1}{2}}
\,.
\end{equation}
is defined up to the phase factor $\eta$. Notice that from Eq.
(\ref{(npnm)}) we understand that the second equation of Eqs.
(\ref{(1)}) gives $\hat f^{(-)}=0$ for $n_{r}=0$.

For $\kappa_{j}=j+1/2=l$ we use the same procedure finding that
$2s_{+}=l+1$, $2s_{-}=l$, $n_{+}=n_{-}=n_{r}$ and
\begin{equation}
\frac{N_{-}}{N_{+}}=\frac{2l+1}{2n_{r}+2k+1}\,.
\end{equation}
In this case the normalized radial wave functions are
\begin{eqnarray}
\hat f^{(+)}(r)_{|\kappa_{j}=j+1/2}&=&
N\left[\frac{n_{r}+k+\frac{1}{2}}{n_{r}+l+\frac{1}{2}}\right]^{\frac{1}{2}}
\nonumber\\
&&\times \sin^{l+1}\omega r \cos^{k}\omega r
P_{n_{r}}^{(l+\frac{1}{2},k-\frac{1}{2})}(\cos 2\omega r)\,,
\label{(2)}\\
\hat f^{(-)}(r)_{|\kappa_{j}=j+1/2}&=& N
\left[\frac{n_{r}+l+\frac{1}{2}}{n_{r}+k+\frac{1}{2}}\right]^{\frac{1}{2}}
\nonumber\\
&&\times\sin^{l}\omega r \cos^{k+1}\omega r
P_{n_{r}}^{(l-\frac{1}{2},k+\frac{1}{2})}(\cos 2\omega r)\,.
\nonumber
\end{eqnarray}

The energy levels result from Eq. (\ref{(quant)}). Bearing in mind
that $\omega k= m$ and $\omega\epsilon=E-\omega/2$, and defining
the {\em principal} quantum number $n=2n_{r}+l$ we obtain the
discrete energy spectrum \cite{COTA}
\begin{equation}\label{(enlev)}
E_{n}=m
+\omega\left(n+\frac{3}{2}\right)\,,\quad n=0,1,2,....
\end{equation}
These levels are degenerated. For a given $n$ our auxiliary
quantum number $l$ takes either all the odd values from $1$ to
$n$, if $n$ is odd, or the even values from $0$ to $n$, if $n$ is
even. In both cases  we have $j=l\pm 1/2$ for each $l$, which
means that $j=1/2,3/2,...,n+1/2$. The selection rule for
$\kappa_{j}$ is more complicated since it is determined by both
the quantum numbers  $n$ and $j$. If $n$ is even then the even
$\kappa_{j}$ are positive while the odd $\kappa_{j}$ are negative.
For odd $n$ we are in the opposite situation, with odd positive or
even negative values of $\kappa_{j}$. Thus it is clear that for
each given pair $(n,j)$ we have only one value of $\kappa_{j}$.
With these specifications and by taking into account that for each
$j$ we have $2j+1$ different values of $m_{j}$, we can conclude
that the degree of degeneracy of the level $E_{n}$ is
$(n+1)(n+2)$.

Since the solutions (\ref{(1)}) and (\ref{(2)}) are completely
determined by the values of $n$ and $j$, we denote by $\hat
f^{(\pm)}_{n,j}$ the radial wave functions (\ref{(1)}) and
(\ref{(2)}). With their help we can write the functions
$f^{(\pm)}_{n,j}$ (i.e. the components of ${\cal F}$)  using the
inverse of the transformation (\ref{(uder)}). Thus from Eq.
(\ref{(psol)}) we find the definitive form of the normalized
particle-like energy eigenspinors of the regular modes,
\begin{eqnarray}\label{(defu)}
U_{n,j,m_{j}}({x})&=&\omega \csc \omega r \cos^{3/2}\omega r\nonumber\\
&&\times\left[ \left(\cos\frac{\omega r}{2}\hat
f^{(+)}_{n,j}(r)+\sin\frac{\omega r}{2}
\hat f^{(-)}_{n,j}(r)\right)\Phi^{+}_{m_{j},\kappa_{j}}(\theta,\phi)\right.\\
&&+\left.\left(-\sin\frac{\omega r}{2}\hat f^{(+)}_{n,j}(r)+
\cos\frac{\omega r}{2} \hat
f^{(-)}_{n,j}(r)\right)\Phi^{-}_{m_{j},\kappa_{j}}(\theta,\phi)\right]e^{-iE_n
t}. \nonumber
\end{eqnarray}
The antiparticle-like energy eigenspinors can be derived directly
by using the charge conjugation \cite{BJD}. These are
\begin{equation}
V_{n,j,m_{j}}=(U_{n,j,m_{j}})^{c}\equiv C
(\overline{U}_{n,j,m_{j}})^{T} \,,\quad C=i\gamma^{2}\gamma^{0}\,.
\end{equation}
Furthermore, we can verify that all these normalized energy
eigenspinors have good orthogonality properties obeying
\begin{eqnarray}
&&\left<{U}_{n,j,m_{j}},U_{n',j',m_{j}'}\right>=\left<{V}_{n,j,m_{j}},
 V_{n',j',m_{j}'}\right>
=\delta_{n,n'}\delta_{j,j'}\delta_{m_{j},m_{j}'}\,,\\
&&\left<{U}_{n,j,m_{j}},V_{n',j',m_{j}'}\right>=\left<{V}_{n,j,m_{j}},
 U_{n',j',m_{j}'}\right>
=0\,,
\end{eqnarray}
where the scalar product (\ref{sp}) is calculated in the chart
$\{t,r,\theta,\phi\}$ using the weight function (\ref{(mus)}) that
now reads
\begin{equation}\label{(muss)}
\mu(r,\theta)=\frac{w(r)^3}{v(r)^2}r^2\sin(\theta)=\frac{1}{\omega^{2}
} \sin^{2}\omega r\, {\rm sec}^{3}\omega r \sin(\theta)\,,
\end{equation}
and taking into account that the angular spinors are normalized with
respect to the angular scalar product \cite{TH}. We observe that the
factors $\omega$ and $1/\omega^2$ can be removed simultaneously from
Eqs. (\ref{(defu)}) and respectively (\ref{(mu)}).

The final result is that for $m\ge\omega/2$, when only regular
modes are allowed, the quantum Dirac field on CAdS reads
\begin{equation}
\psi({x})=\sum_{n,j,m_{j}}\left[U_{n,j,m_{j}}({x})
a_{n,j,m_{j}}+V_{n,j,m_{j}}({x})b^{\dagger}_{n,j,m_{j}} \right]\,.
\end{equation}
This field can be quantized supposing that the particle ($a$,
$a^{\dagger}$) and antiparticle ($b$, $b^{\dagger}$) operators
satisfy usual anticommutation relations from which the
non-vanishing ones are
\begin{equation}
\{a_{n,j,m_{j}}, a_{n',j',m_{j}'}^{\dagger}\}= \{b_{n,j,m_{j}},
b_{n',j',m_{j}'}^{\dagger}\} =\delta_{n,n'}\delta_{j,j'}
\delta_{m_{j},m_{j}'} {\bf 1}\,,
\end{equation}
where ${\bf 1}$ is the identity operator. Then the one-particle
operators obtained via Noether theorem, using Eq. (\ref{opo}) have
correct forms and satisfy the conditions (\ref{algXX}). After a
little calculation we obtain the Hamiltonian operator
\begin{equation}
{\bf H}=\sum_{n,j,m_{j}}E_n\left[a_{n,j,m_{j}}^{\dagger}
a_{n,j,m_{j}}+b^{\dagger}_{n,j,m_{j}} b_{n,j,m_{j}}\right]\,,
\end{equation}
while the charge operator reads
\begin{equation}
{\bf Q}=\sum_{n,j,m_{j}}\left[a_{n,j,m_{j}}^{\dagger}
a_{n,j,m_{j}}-b^{\dagger}_{n,j,m_{j}} b_{n,j,m_{j}}\right]\,.
\end{equation}
The angular operators  $\vec{\bf J}^2$, ${\bf J}_3$ and ${\bf K}$,
which are diagonal in this basis, take the same form as ${\bf H}$
but depending on their own  eigenvalues $j(j+1)$, $m_j$ and
$\kappa_j$ respectively. Thus, we conclude that the quantum modes of
the Dirac fermions on AdS backgrounds we derived here are determined
by the complete set of commuting operators  $\{ {\bf H},{\bf
Q},{\vec{\bf J}}^2,{\bf K}, {\bf J}_3\}$.  All these operators have
similar structures and properties like those of the usual quantum
field theory in flat spacetime.

\subsection{Dirac quantum modes in dS spacetime}

The first solution of the Dirac equation on dS backgrounds was
found in Ref.  \cite{OT} based on a diagonal gauge in central
charts. Other interesting solutions in the null tetrad gauge of
these charts were derived recently in Ref.  \cite{LO}. However,
here we restrict ourselves to present only the solutions that can
be obtained in our Cartesian gauge \cite{COTA1}.

Let us consider the problem of the massive Dirac particle on a dS
background of radius $R=1/\omega$ defined by Eq. (\ref{hip}) for
$\epsilon=1$. There exists a special central chart
$\{t,r,\theta,\phi\}$ with the line element
\begin{equation}\label{(le)}
ds^{2}=\frac{1}{\cosh^{2}\omega r}
\left[dt^{2}-dr^{2}-\frac{1}{\omega^{2}} \sinh^{2}\omega r~
(d\theta^{2}+\sin^{2}\theta~d\phi^{2})\right]\,,
\end{equation}
from which  we can identify the functions $v$ and $w$ we need for
writing down the radial Hamiltonian,
\begin{equation}
H_r=\begin{array}{|cc|}
  \frac{\textstyle\omega k}{\textstyle\cosh \omega r}& -\frac{\textstyle d}{\textstyle dr}+
\frac{\textstyle\omega\kappa_{j}}{\textstyle\sinh \omega r}\\
&\\
\frac{\textstyle d}{\textstyle
dr}+\frac{\textstyle\omega\kappa_{j}} {\textstyle\sinh\omega r} &
-\frac{\textstyle\omega k}{\textstyle\cosh \omega r}
\end{array}\,,
\end{equation}
with the same notation, $k=m/\omega$, as in the previous case.
This is the basic piece of the radial problem which must determine
the radial functions $f^{(\pm)}$ of the solutions (\ref{(psol)}).

This radial problem has a hidden supersymmetry like in the AdS
case \cite{COTA1}. This can be pointed out with the help of the
transformation ${\cal F}\to \hat{\cal F}=U(r){\cal F}$ where
${\cal F}=(f^{(+)},f^{(-)})^T$, $\hat{\cal F}=(\hat f^{(+)},\hat
f^{(-)})^T$ and
\begin{equation}\label{(uder)}
U(r)=\begin{array}{|cc|}
    \cosh \frac{\textstyle \omega r}{\textstyle 2}&-i\sinh
\frac{\textstyle \omega r}{\textstyle 2}\\
&\\
    i\sinh \frac{\textstyle \omega r}{\textstyle 2}&\cosh
\frac{\textstyle \omega r}{\textstyle 2}
\end{array}\,.
\end{equation}
A little calculation shows us that  the transformed  Hamiltonian,
\begin{equation}\label{(newh)}
\hat H_r =U(r)H_rU^{-1}(r)-i\frac{\omega}{2} 1_{2\times 2},
\end{equation}
which gives the new eigenvalue problem
\begin{equation}\label{(trrp)}
\hat H_r \hat{\cal F}=\left(E-i\frac{\omega}{2}\right)\hat{\cal
F},
\end{equation}
has supersymmetry since it  has  the specific form (\ref{(ssh)})
where   $\nu=\omega(k-i\kappa_{j})$  and
\begin{equation}\label{(super)}
W(r)=\omega(ik\tanh\omega r + \kappa_{j}\coth \omega r)
\end{equation}
is the superpotential of the radial problem \cite{COTA1}. The
transformed radial wave functions  $\hat f^{(\pm)}$  satisfy the
second order equations resulted from the square of Eq.
(\ref{(trrp)}). These are
\begin{equation}\label{(2e)}
\left(-\frac{d^2}{dr^2}-\omega^{2}\frac{ik(ik\pm
1)}{\cosh^{2}\omega r}+ \omega^{2}\frac{\kappa_{j}(\kappa_{j}\pm
1)}{\sinh^{2}\omega r}\right) \hat f^{(\pm)}(r)=
\omega^{2}\epsilon^{2}\hat f^{(\pm)}(r)\label{(od1)},\\
\end{equation}
where we have denoted $\epsilon=E/\omega-i/2$.

The solutions of Eqs. (\ref{(2e)})  are well-known to be expressed
in terms of Gauss  hypergeometric functions \cite{AS},
$F_{\pm}(y)\equiv F(\alpha_{\pm},\beta_{\pm},\gamma_{\pm},y)$,
depending on the new variable $y=-\sinh^{2}\omega r$, as
\begin{equation}\label{(gsol)}
\hat f^{(\pm)}(y)=N_{\pm}(1-y)^{p_{\pm}}y^{s_{\pm}} F_{\pm}(y)
\end{equation}
where
\begin{equation}
\alpha_{\pm}=s_{\pm}+p_{\pm}+\frac{i\epsilon}{2},\quad
\beta_{\pm}=s_{\pm}+p_{\pm}-\frac{i\epsilon}{2},\quad
\gamma_{\pm}=2s_{\pm}+\frac{1}{2},
\end{equation}
$N_{\pm}$ are normalization factors while the parameters
$p_{\pm}$ and $s_{\pm}$ are related with  $k$ and $\kappa_{j}$
through
\begin{eqnarray}
2s_{\pm}(2s_{\pm}-1)&=&\kappa_{j}(\kappa_{j}\pm 1),\label{(s)}\\
2p_{\pm}(2p_{\pm}-1)&=&ik(ik\pm 1).\label{(p)}
\end{eqnarray}

Furthermore, we have to find the suitable values of these
parameters such that the functions (\ref{(gsol)}) should be
solutions of the transformed radial problem (\ref{(trrp)}). If we
replace (\ref{(gsol)}) in (\ref{(trrp)}), after a few
manipulation, we obtain
\begin{eqnarray}
&&y(1-y)\frac{dF_{\pm}(y)}{dy}-y\left(p_{\pm}\pm\frac{ik}{2}\right)F_{\pm}(y)
+(1-y)\left(s_{\pm}\pm\frac{\kappa_{j}}{2}\right)F_{\pm}(y)\nonumber\\
&&=\frac{\eta}{2}\frac{N_{\mp}}{N_{\pm}}\left(\kappa_{j}\pm\frac{1}{2}+
i\frac{M\pm
E}{\omega}\right)y^{s_{\mp}-s_{\pm}+1/2}(1-y)^{p_{\mp}-p_{\pm}
+1/2}F_{\mp}(y),\label{(idf)}
\end{eqnarray}
where $\eta=\pm 1$. These equations are nothing other than the usual
identities of  hypergeometric functions  if the values of $s_{\pm}$,
$p_{\pm}$ and $N_{+}/N_{-}$ are correctly matched. First we observe
that the differences $s_{+}-s_{-}$ and $p_{+}-p_{-}$ must be
half-integer  since we work with analytic functions of $y$. This
means that the allowed groups of solutions of (\ref{(s)}) are
\begin{eqnarray}
2s_{+}^{1}=-\kappa_{j}~~~~~ ,&\quad&  2s_{+}^{2}=\kappa_{j}+1,\nonumber\\
2s_{-}^{1}=-\kappa_{j}+1 ,&\quad&
2s_{-}^{2}=\kappa_{j},\label{(kka)}
\end{eqnarray}
while  Eq. (\ref{(p)}) gives us
\begin{eqnarray}
2p_{+}^{1}=-ik~~~~~ ,&\quad&  2p_{+}^{2}=ik+1,\nonumber\\
2p_{-}^{1}=-ik+1 ,&\quad&  2p_{-}^{2}=ik.\label{(ppa)}
\end{eqnarray}
On the other hand, it is known that the hypergeometric functions
of (\ref{(gsol)}) are analytical on the domain
$D_{y}=(-\infty,0]$, corresponding to $D_{r}$, only if
$\Re(\gamma_{\pm})>\Re(\beta_{\pm})>0$ \cite{AS}. Moreover, their
factors must be  regular on this domain including  $y=0$.
Obviously, both these conditions are accomplished if we take
$s_{\pm}> 0$. We specify that there are no restrictions upon the
values of the parameters $p_{\pm}$.

We have hence all the possible combinations of parameter values
giving the solutions of the second order equations (\ref{(2e)})
which satisfy the transformed radial problem. These solutions will
be denoted  by $(a,b)$, $a,b=1,2$, understanding that the
corresponding parameters are $s_{\pm}^{a}$, $p_{\pm}^{b}$, as
given by  (\ref{(kka)}) and (\ref{(ppa)}), and
\begin{equation}
\alpha_{\pm}^{(a,b)}=s_{\pm}^{a}+p_{\pm}^{b}+\frac{i\epsilon}{2},\quad
\beta_{\pm}^{(a,b)}=s_{\pm}^{a}+p_{\pm}^{b}-\frac{i\epsilon}{2},
\quad \gamma_{\pm}^{(a)}=2s_{\pm}^{a}+\frac{1}{2}.
\end{equation}
The condition  $s_{\pm}> 0$ requires to chose $a=1$ when
$\kappa_{j}=-j-1/2$, and $a=2$  if $\kappa=j+1/2$. Thus, for each
given set $(E,j,\kappa_{j})$ we have a pair of different radial
solutions, with $b=1,2$. Therefore, it is convenient to denote the
transformed radial wave functions (\ref{(gsol)}) by $\hat
f^{(\pm)}_{E,\kappa_j,a,b}$, bearing in mind that the value of $a$
determines that of $\kappa_{j}$. The last step is to calculate the
values of $N_{+}/N_{-}$. From (\ref{(idf)}) it results
\begin{eqnarray}
\kappa_{j}=-j-\frac{1}{2}:
&&\eta\frac{N_{-}^{(1,1)}}{N_{+}^{(1,1)}}=
-\frac{\alpha_{+}^{(1,1)}}{\gamma_{+}^{(1)}}, \qquad
~~~\eta\frac{N_{-}^{(1,2)}}{N_{+}^{(1,2)}}=
\frac{\beta_{+}^{(1,2)}}{\gamma_{+}^{(1)}}-1, \\
\kappa_{j}=j+\frac{1}{2}:
&&\eta\frac{N_{+}^{(2,1)}}{N_{-}^{(2,1)}}=
1-\frac{\beta_{-}^{(2,1)}}{\gamma_{-}^{(2)}}, \qquad
\eta\frac{N_{+}^{(2,2)}}{N_{-}^{(2,2)}}=
\frac{\alpha_{-}^{(2,2)}}{\gamma_{-}^{(2)}}.
\end{eqnarray}
Notice that here  $\gamma_{+}^{(1)}= \gamma_{-}^{(2)}=j+1$.

Now we can restore the form of the original radial wave functions
of (\ref{(psol)}) using the inverse of the transformation
(\ref{(uder)}). In our new notation these wave functions are
\cite{COTA1}
\begin{equation}\label{(ff)}
f^{(\pm)}_{E,\kappa_j,a,b}(r)=\cosh\frac{\omega r}{2}\hat
f^{(\pm)}_{E,\kappa_j,a,b}(r) \pm i\sinh\frac{\omega r}{2}\hat
f^{(\mp)}_{E,\kappa_j,a,b}(r).
\end{equation}
For very large $r$ (when $y\to -\infty$) the hypergeometric
functions behave as $F_{\pm}(y)\sim (-y)^{-\alpha_{\pm}}$
\cite{AS}. Thereby we deduce that $\hat
f^{(\pm)}_{E,\kappa_j,a,b}\sim \exp(-i\epsilon \omega r)$ and
\begin{equation}\label{(ffa)}
f^{(\pm)}_{E,\kappa_j,a,b}\sim e^{-iEr}.
\end{equation}
This means that the functions (\ref{(ff)}) represent tempered
distributions corresponding to a continuous energy spectrum. On
the other hand, Eq. (\ref{(ffa)}) indicates that this  energy
spectrum covers  the whole real axis, as it seems to be natural
since the metric is not asymptotically flat.  Anyway, it is clear
that the energy spectrum is continuous, without discrete part,
while the energy levels are infinitely degenerated since there are
no restrictions upon the values of $j$, which can be any positive
half-integer.

In these conditions, the fundamental solutions can not be normalized
in the generalized sense and, therefore, the quantization the Dirac
field can not be done in this chart. For doing this we are forced to
consider moving charts.

\section{Dirac fermions in moving frames of the dS spacetime}

The first solutions of the Dirac equation in moving charts  with
spherical coordinates of the dS manifolds were derived in Ref.
\cite{SHI} and normalized in Ref. \cite{C4}.  Other solutions of
this equation in moving frames with Cartesian coordinates have been
reported \cite{BFGF} but these solutions are not correctly
normalized. We have constructed the normalized solutions in this
framework using the helicity basis in momentum representation
\cite{C3}. Our theory of external symmetry offered us the operators
we need for deriving the normalized fundamental solutions and
quantizing the Dirac field in usual manner.

\subsection{Observables in dS spacetime}

We remain in the case of the dS spacetime, $M$, of radius
$R=1/\omega$, for which we keep the definition (\ref{hip}) and the
previous notations. As mentioned, this manifold is the homogeneous
space of the pseudo-orthogonal group $SO(4,1)$. This group is in the
same time the  gauge group of the metric $\eta(1)=(1,-1,-1,-1,-1)$
and the isometry group, $I(M)$, of the dS spacetime. For this reason
it is convenient to use the covariant real parameters
$\xi^{AB}=-\xi^{BA}$ since in this case the orbital basis-generators
of the representation of $SO(4,1)$, carried by the space of the
scalar functions over $M^{5}$, have the standard form (\ref{LAB5}).
They will give us directly the orbital basis-generators $L_{(AB)}$
of the scalar representations of $I(M)$. Indeed, starting with the
functions $Z^{A}(x)$ that solve the equation (\ref{hip}) in a given
chart $\{x\}$, one can write down the operators (\ref{LAB5}) in the
form (\ref{La}), finding thus the generators $L_{(AB)}$ and
implicitly the components $k^{\mu}_{(AB)}(x)$ of the Killing vectors
associated to the parameters $\xi^{AB}$ \cite{C2}. Furthermore, one
has to calculate the spin parts $S_{(AB)}$, according to Eqs.
(\ref{Sx}) and (\ref{Om}), arriving to the final form of the
basis-generators $X_{(AB)}=L_{(AB)}+S_{(AB)}$ of the spinor
representation of $S(M)$.

In the dS spacetime there are many moving frames of physical
interest. First we consider the moving  local chart
$\{t,r,\theta,\phi\}_*$ associated to the Cartesian one $\{t,
\vec{x}\}_*$ with the line element
\begin{equation}\label{mssu}
ds^{2}=dt^2 - e^{2\omega t} d\vec{x}^2\,.
\end{equation}
The time $t\in (-\infty, \infty)$ of this chart is interpreted as
the {\em proper} time of an observer at $\vec{x}=0$. Another moving
chart which play here a special role is the chart $\{t_c,
\vec{x}\}_*$ with the {\em conformal} time $t_c$ and Cartesian
spaces coordinates $x^i$ defined by
\begin{eqnarray}
Z^0&=&-\frac{1}{2\omega^2 t_c}\left[1-\omega^2({t_c}^2 -
r^2)\right]
\nonumber\\
Z^5&=&-\frac{1}{2\omega^2 t_c}\left[1+\omega^2({t_c}^2 -
r^2)\right]
\label{Zx}\\
Z^i&=&-\frac{1}{\omega t_c}x^i \nonumber
\end{eqnarray}
with $r=|\vec{x}|$. This chart covers only a half of the manifold
$M$, for $t_c\in (-\infty,\,0)$  and $\vec{x}\in D\equiv {\Bbb
R}^3$. Nevertheless, it has the advantage of a simple conformal
flat line element \cite{BD},
\begin{equation}\label{mconf}
ds^{2}=\frac{1}{\omega^2
{t_c}^2}\left({dt_c}^{2}-d\vec{x}^2\right)\,.
\end{equation}
Moreover, the conformal time $t_c$ is related to the proper time
$t$ through
\begin{equation}
\omega t_c =-e^{-\omega t}\,.
\end{equation}

In what follows we study the Dirac field in the chart
$\{t,\vec{x}\}_*$ using the conformal time as a helpful auxiliary
ingredient. The form of the line element (\ref{mconf}) allows one to
choose the simple Cartesian gauge with the non-vanishing tetrad
components \cite{SHI}
\begin{equation}\label{tt}
e^{0}_{0}=-\omega t_{c}\,, \quad e^{i}_{j}=-\delta^{i}_{j}\,\omega
t_c \,,\quad \hat e^{0}_{0}=-\frac{1}{\omega t_{c}}\,, \quad \hat
e^{i}_{j}=-\delta^{i}_{j}\, \frac{1}{\omega t_c}\,.
\end{equation}
In this gauge the Dirac operator reads
\begin{eqnarray}\label{ED1}
E_D&=&-i\omega
t_c\left(\gamma^0\partial_{t_{c}}+\gamma^i\partial_i\right)
+\frac{3i\omega}{2}\gamma^{0} \nonumber\\
&=&i\gamma^0\partial_{t}+ie^{-\omega t}\gamma^i\partial_i
+\frac{3i\omega}{2}\gamma^{0}
\end{eqnarray}
and the weight function of the scalar product (\ref{sp}) is
\begin{equation}\label{mu}
\mu=(-\omega t_{c})^{-3}=e^{3\omega t}\,.
\end{equation}

The next step is to calculate the basis-generators $X_{(AB)}$ of the
spinor representation of $S(M)$ in this gauge since these are the
main operators that commute with $E_{D}$. The group $SO(4,1)$
includes the subgroup $E(3)=T(3)\circledS SO(3)$ which is just the
isometry group of the 3-dimensional Euclidean space of our moving
charts, $\{t_c,\vec{x}\}_*$ and $\{t,\vec{x}\}_*$, formed by ${\Bbb
R}^3$ translations, $x^i\to x^i+a^i$, and proper rotations, $x^i\to
R^{i\,\cdot}_{\cdot\,j} x^j$ with $R\in SO(3)$ \cite{W}. Therefore,
the basis-generators of its universal covering group, $\tilde
E(3)=T(3)\circledS SU(2)\subset S(M)$, can be interpreted as the
components of the momentum, $\vec{P}$, and total angular momentum,
$\vec{J}$, operators. The problem of the Hamiltonian operator seems
to be more complicated, but we know that in the mentioned static
central charts with the static time $t_{s}$ this is $H=\omega
X_{(05)}=i\partial_{t_{s}}$ \cite{C2}. Thus the Hamiltonian operator
and the components of the momentum and total angular momentum
operators  ($P^i$ and $J^i=\varepsilon_{ijk}J_{jk}/2$ respectively)
can be identified as being the following basis-generators of $S(M)$
\begin{eqnarray}
H&\equiv&\omega
X_{(05)}=-i\omega(t_{c}\partial_{t_{c}}+x^{i}\partial_i)
\label{Gi}\\
P^{i}&\equiv&\omega\left(X_{(5i)}-X_{(0i)}\right)=-i\partial_{i}\label{Gip}\\
J_{ij}&\equiv&X_{(ij)}=-i(x^i\partial_j-x^j\partial_i)+S_{ij}
\end{eqnarray}
after which one remains with the three basis-generators
\begin{equation}\label{Gf}
N^i\equiv X_{(5i)}+X_{(0i)}=\omega ({t_{c}}^2- r^2)P^i + 2 x^i
H+2\omega( S_{i0}t_{c}+ S_{ij}x^j)\,,
\end{equation}
which do not have an immediate physical significance. The $SO(4,1)$
transformations corresponding to these basis-generators and the
associated isometries of the chart $\{t_{c}, \vec{x}\}_*$ are
briefly presented in Appendix A.

In the other local chart, $\{t,\vec{x}\}_*$, we have the same
operators $\vec{P}$ and $\vec{J}=\vec{L}+\vec{S}$ (with
$\vec{L}=\vec{x}\times \vec{P}$) whose  components are the $\tilde
E(3)$ generators,  while the Hamiltonian operator takes the form
\begin{equation}\label{Ham}
H=i\partial_{t}+\omega\, \vec{x}\cdot\vec{P}\,,
\end{equation}
where  the second term, due to the external gravitational field,
leads to the commutation rules
\begin{equation}\label{cHP}
[H,\,P^i]=i\omega P^i \,.
\end{equation}
We observe that in this chart the operators $K^i\equiv X_{(0i)}$
are the analogous of the basis-generators of the Lorentz boosts of
$SL(2,\Bbb C)$ since in the limit of $\omega\to 0$, when
$(\ref{mssu})$ equals the Minkowski line element, the operators
$H=P^0,\,P^i,\,J^i$ and $K^i$ become the generators of the spinor
representation of the group $T(4)\circledS SL(2,\Bbb C)$ (i.e. the
universal covering group of the Poincar\' e group \cite{W,TH}).

In both the charts we used here the generators
(\ref{Gi})-(\ref{Gf}) are self-adjoint with respect to the scalar
product (\ref{sp}) with the weight function (\ref{mu}) if we
consider the usual boundary conditions on $D\equiv {\Bbb R}^3$.
Therefore, for any generator $X$ we have
$\left<X\psi,\psi'\right>=\left<\psi,X\psi'\right>$ if (and only
if) $\psi$ and $\psi'$ are solutions of the Dirac equation which
behave as  tempered distributions or square integrable spinors
with respect to the scalar product (\ref{sp}). Moreover, all these
generators commute with the Dirac operator $E_D$.

\subsection{Polarized plane wave solutions}

The plane wave solutions of the Dirac equation with $m\not =0$ must
be eigenspinors of the momentum operators $P^i$ corresponding to the
eigenvalues $p^i$, with the same time modulation as the spherical
waves. Therefore, we have to look for particular solutions in the
chart $\{t_c,\vec{x}\}_*$ involving either positive or negative
frequency plane waves. Bearing in mind that these must be related
among themselves through the charge conjugation, we assume that, in
the standard representation of the Dirac matrices (with diagonal
$\gamma^0$ \cite{TH}), they have the form
\begin{eqnarray}
\psi^{(+)}_{\vec{p}}&=&  \left(
\begin{array}{c}
f^{+}(t_{c}) \,\alpha(\vec{p})\\
g^{+}(t_{c})\, \frac{\textstyle
\vec{\sigma}\cdot\vec{p}}{\textstyle p} \,\alpha(\vec{p})
\end{array}\right)
e^{i\vec{p}\cdot\vec{x}} \label{psi+}\\
\psi^{(-)}_{\vec{p}}&=&  \left(
\begin{array}{c}
g^{-}(t_{c})\, \frac{\textstyle
\vec{\sigma}\cdot\vec{p}}{\textstyle p}
\,\beta(\vec{p})\\
f^{-}(t_{c}) \,\beta(\vec{p})
\end{array}\right)
e^{-i\vec{p}\cdot\vec{x}}\label{psi-}
\end{eqnarray}
where $p=|\vec{p}|$, $\sigma_i$ denotes the Pauli matrices while
$\alpha$ and $\beta$ are arbitrary Pauli spinors depending on
$\vec{p}$. Replacing these spinors in the Dirac equation given by
(\ref{ED1}) and denoting $k=m/\omega$ and
$\nu_{\pm}=\frac{1}{2}\pm ik$, we find  equations of the form
(\ref{H2}) whose solutions can be written in terms of Hankel
functions as
\begin{eqnarray}
&&f^{+}= (-f^{-})^{*}=C{t_{c}}^{2}e^{\pi k/2}H^{(1)}_{\nu_{-}}(-p
t_{c})
 \label{fg1}\\
&&g^{+}= (-g^{-})^{*}=C{t_{c}}^{2} e^{-\pi
k/2}H^{(1)}_{\nu_{+}}(-p t_{c})\,. \label{fg2}
\end{eqnarray}
The integration constant $C$ will be calculated from the
orthonormalization condition  in the momentum scale.

The plane wave solutions are determined up to the significance of
the Pauli spinors $\alpha$ and $\beta$. For $\vec{p}\not = 0$
these can be treated as in the flat case \cite{SW1,TH} since, in
the tetrad gauge (\ref{tt}), the spaces of these spinors carry
unitary linear representations of the $\tilde E(3)$ group. Indeed,
a transformation (\ref{es}) produced by $(A, \phi_{A,\vec{a}})\in
\tilde E(3)\subset S(M)$ where $A\in SU(2)$ and $\vec{a}\in {\Bbb
R}^3$ involves the usual linear isometry of $E(3)$, $x^i\to
x^{\prime\,i}=\phi^{i}_{A,\vec{a}}(\vec{x})\equiv
\Lambda^{i\,\cdot}_{\cdot\,j}(A) x^j+a^i$  with $\Lambda(A)\in
SO(3)$, and the global transformation $\psi(t,\vec{x})\to
\psi^{\prime}(t,{\vec{x}}^{\,\prime})= \rho(A)\psi(t,\vec{x})$.
Consequently, the Pauli spinors transform according to the
unitary (linear) representation
\begin{equation}\label{tri}
\alpha(\vec{p})\to
e^{-i\vec{a}\cdot\vec{p}}\,A\,\alpha[\Lambda(A)^{-1} \vec{p}\,]
\end{equation}
(and similarly for $\beta$) that preserves orthogonality. This
means that any pair of orthogonal spinors
$\tilde\xi_{\sigma}(\vec{p})$ with polarizations  $\sigma=\pm 1/2$
(obeying $\tilde \xi_{\sigma}^{+}\tilde \xi_{\sigma
'}=\delta_{\sigma\sigma'}$) represents a good basis in the space
of Pauli spinors
\begin{equation}\label{alpha}
\alpha(\vec{p})= \sum_{\sigma} \tilde \xi_{\sigma}(\vec{p})
a(\vec{p},\sigma)
\end{equation}
whose components, $a(\vec{p}, \sigma)$, are the particle wave
functions in momentum representation. According to the standard
interpretation of the negative frequency terms \cite{SW1,TH}, the
corresponding basis of the space of $\beta$ spinors is defined by
\begin{equation} \label{beta}
\beta(\vec{p})= \sum_{\sigma} \tilde \eta_{\sigma}(\vec{p})
[b(\vec{p},\sigma)]^{*} \,, \quad \tilde \eta_{\sigma}(\vec{p})=
i\sigma_2 [\tilde \xi_{\sigma}(\vec{p})]^{*}
\end{equation}
where $b(\vec{p},\sigma)$ are the antiparticle wave functions. It
remains to choose a specific basis, using supplementary physical
assumptions. Since it is not sure that the so called spin basis
\cite{SW1} can be correctly defined in the dS geometry, we prefer
the {\em helicity} basis. This is formed by the orthogonal Pauli
spinors of helicity $\lambda=\pm 1/2$ which fulfill
\begin{equation}\label{heli}
\vec{\sigma}\cdot\vec{p}\,\,\tilde\xi_{\lambda}(\vec{p})=2p\lambda\,
\tilde\xi_{\lambda}(\vec{p}) \,, \quad
\vec{\sigma}\cdot\vec{p}\,\,\tilde\eta_{\lambda}(\vec{p})=-2p\lambda\,
\tilde\eta_{\lambda}(\vec{p})\,.
\end{equation}

The desired particular solutions of the Dirac equation with
$m\not=0$ result from our starting formulas (\ref{psi+}) and
(\ref{psi-}) where we insert the functions (\ref{fg1}) and
(\ref{fg2}) and the spinors (\ref{alpha}) and (\ref{beta}) written
in the helicity basis (\ref{heli}). It remains to calculate the
normalization constant $C$ with respect to the scalar product
(\ref{sp}) with the weight function (\ref{mu}). After a few
manipulation, in the chart $\{t,\vec{x}\}_*$, it turns out the final
form of the fundamental spinor solutions of positive and negative
frequencies with momentum $\vec{p}$ and helicity $\lambda$,
\begin{eqnarray}
U_{\vec{p},\lambda}(t,\vec{x})&=& i N\left(
\begin{array}{c}
\frac{1}{2}\,e^{\pi k/2}H^{(1)}_{\nu_{-}}(qe^{-\omega t}) \,
\tilde\xi_{\lambda}(\vec{p})\\
\lambda\, e^{-\pi k/2}H^{(1)}_{\nu_{+}}(qe^{-\omega t})
\,\tilde\xi_{\lambda}(\vec{p})
\end{array}\right)
e^{i\vec{p}\cdot\vec{x}-2\omega t}\label{Ups}\\
V_{\vec{p},\lambda}(t,\vec{x})&=&iN  \left(
\begin{array}{c}
-\lambda\,e^{-\pi k/2}H^{(2)}_{\nu_{-}}(qe^{-\omega t})\,
\tilde\eta_{\lambda}(\vec{p})\\
\frac{1}{2}\,e^{\pi k/2}H^{(2)}_{\nu_{+}}(qe^{-\omega t})
\,\tilde\eta_{\lambda}(\vec{p})
\end{array}\right)
e^{-i\vec{p}\cdot\vec{x}-2\omega t}\,,\label{Vps}
\end{eqnarray}
where we introduced the new parameter $q=p/\omega$ and
\begin{equation}
N=\frac{1}{(2\pi)^{3/2}}\sqrt{\pi q}\,.
\end{equation}
Using Eqs. (\ref{H1}) and (\ref{H3}), it is not hard to verify
that these spinors are charge-conjugated to each other,
\begin{equation}\label{conj}
V_{\vec{p},\lambda}=(U_{\vec{p},\lambda})^{c}={\cal C}
(\overline{U}_{\vec{p},\lambda})^T \,, \quad {\cal
C}=i\gamma^2\gamma^0\,,
\end{equation}
satisfy the orthonormalization relations
\begin{eqnarray}
&&\left<U_{\vec{p},\lambda},U_{\vec{p}^{\,\prime},\lambda^{\prime}}\right>=
\left<V_{\vec{p},\lambda},V_{\vec{p}^{\,\prime},\lambda^{\prime}}\right>=
\delta_{\lambda\lambda^{\prime}}\delta^3
(\vec{p}-\vec{p}^{\,\prime})\,,
\label{orto1}\\
&&\left<U_{\vec{p},\lambda},V_{\vec{p}^{\,\prime},\lambda^{\prime}}\right>=
\left<V_{\vec{p},\lambda},U_{\vec{p}^{\,\prime},\lambda^{\prime}}\right>=
0\,,\label{orto2}
\end{eqnarray}
and represent a {\em complete} system of solutions in the sense
that
\begin{equation}\label{compl}
\int d^3 p \sum_{\lambda}\left[
U_{\vec{p},\lambda}(t,\vec{x})U^{+}_{\vec{p},\lambda}(t,\vec{x}^{\,\prime})+
V_{\vec{p},\lambda}(t,\vec{x})V^{+}_{\vec{p},\lambda}(t,\vec{x}^{\,\prime})
\right]=e^{-3\omega t}\delta^3 (\vec{x}-\vec{x}^{\,\prime})\,.
\end{equation}
Let us observe that the factor $e^{-3\omega t}$ is exactly the
quantity necessary to compensate the weight function (\ref{mu}).
Other important properties are
\begin{eqnarray}
P^i\,U_{\vec{p},\lambda}= p^i\, U_{\vec{p},\lambda}\,, &\quad&
P^i\,V_{\vec{p},\lambda}=-p^i\, V_{\vec{p},\lambda} \,,\label{PUV}\\
W\,U_{\vec{p},\lambda}= p\lambda U_{\vec{p},\lambda}\,, &\quad&
W\,V_{\vec{p},\lambda}=-p\lambda V_{\vec{p},\lambda}\,,\label{WUV}
\end{eqnarray}
where
\begin{equation}\label{PL}
W=\vec{J}\cdot \vec{P}=\vec{S}\cdot\vec{P}
\end{equation}
is the helicity operator which is analogous to the time-like
component of the four-component Pauli-Lubanski operator of the
Poincar\' e algebra \cite{W}. Thus, we arrive at the conclusion that
the fundamental solutions (\ref{Ups}) and (\ref{Vps}) form a
complete system of common eigenspinors of the operators $P^i$ and
$W$. Since the spin was fixed a priori by choosing the
representation $\rho$, we consider that the complete set of
commuting operators which  determines separately each of the sets of
the particle or antiparticle eigenspinors is $\{E_{D},\,P^i,\,W\}$.
Finally we note that these solutions can be redefined at any time
with other momentum dependent phase factors as
\begin{equation}\label{gaugeUV}
U_{\vec{p},\lambda}\to e^{i\chi(\vec{p})}
U_{\vec{p},\lambda}\,,\quad V_{\vec{p},\lambda}\to
e^{-i\chi(\vec{p})} V_{\vec{p},\lambda}\,,\quad \chi(\vec{p})\in
\Bbb R\,,
\end{equation}
without to affect the above properties.

In the case  $m=0$ (when $k=0$) it is convenient to consider the
chiral representation of the Dirac matrices (with diagonal
$\gamma^5$ \cite{SW1}) and the chart $\{t_c,\vec{x}\}_*$. We find
that the fundamental solutions in helicity basis of the left-handed
massless Dirac field,
\begin{eqnarray}
U^0_{\vec{p},\lambda}(t_c,\vec{x})&=& \lim_{k\to 0}
\frac{1-\gamma^5}{2} U_{\vec{p},\lambda}(t_c,\vec{x})
\nonumber\\
&=&\left(\frac{-\omega t_{c}}{2\pi}\right)^{3/2} \left(
\begin{array}{c}
(\frac{1}{2}-\lambda)\tilde\xi_{\lambda}(\vec{p})\\
0
\end{array}\right)
\,e^{-ipt_{c}+i\vec{p}\cdot\vec{x}} \label{n1}\\
V^0_{\vec{p},\lambda}(t_c,\vec{x})&=& \lim_{k\to 0}
\frac{1-\gamma^5}{2}V_{\vec{p},\lambda}(t_c,\vec{x})
\nonumber\\
&=&\left(\frac{-\omega t_{c}}{2\pi}\right)^{3/2} \left(
\begin{array}{c}
(\frac{1}{2}+\lambda)\tilde\eta_{\lambda}(\vec{p})\\
0
\end{array}\right)
\,e^{ipt_{c}-i\vec{p}\cdot\vec{x}}\,,  \label{n2}
\end{eqnarray}
are non-vanishing only for positive frequency and $\lambda=-1/2$
or negative frequency and $\lambda=1/2$, as in Minkowski
spacetime. Obviously, these solutions have similar properties as
(\ref{conj})-(\ref{WUV}).

\subsection{Spherical waves}

Let us consider now the spherical modes \cite{C4} in moving charts
defined by the common eigenspinors of the complete set
$\{E_D,\vec{P}^2, \vec{J}^2,K, J_3\}$ where $\vec{P}^2$ plays the
role of $H_D$ in static frames. Our purpose is to write down the
particular solutions of the Dirac equation defined as eigenspinors
of the above set of commuting operators corresponding to the
eigenvalues $\{m,p^2, j(j+1), -\kappa_j, m_j\}$ where  $p$ is the
value of the {\em scalar momentum}. In addition, we require these
solutions to be normalized with respect to the relativistic scalar
product defined by Eq. (\ref{sp}) with the wight function
(\ref{mu}).

For solving the above eigenvalue problems it is convenient to start
with the chart $\{t_c,r,\theta,\phi\}_*$ looking for particular
solutions of the form
\begin{equation}
U_{p,\kappa_j, m_j}(x)=\frac{(-\omega
t_c)^{\frac{3}{2}}}{r}\left[f^+_{p,\kappa_j}(t_c,r)
\Phi^+_{m_j,\kappa_j}(\theta,\phi)
+f^-_{p,\kappa_j}(t_c,r)\Phi^-_{m_j,\kappa_j}(\theta,\phi)\right]
\end{equation}
Separating the angular variables in the Dirac equation and denoting
by $k=m/\omega$, after a little calculation, we arrive ar the pair
of equations
\begin{equation}\label{rad}
\left(\pm i\partial_{t_c}+\frac{k}{t_c}
\right)f^{\pm}_{p,\kappa_j}(t_c,r)= \left( -\partial_r\pm
\frac{\kappa_j}{r}\right)f^{\mp}_{p,\kappa_j}(t_c,r)\,.
\end{equation}
In addition,  the eigenvalue problem of $\vec{P}^2$ leads to the
supplemental radial equations
\begin{equation}\label{PP}
\left[-\partial_r^2+\frac{\kappa_j(\kappa_j\pm
1)}{r^2}\right]f^{\pm}_{p,\kappa_j}(t_c,r)=p^2
f^{\pm}_{p,\kappa_j}(t_c,r)\,,
\end{equation}
since the spinors $\Phi^{\pm}_{m_j,\kappa_j}$ are eigenfunctions of
$\vec{L}^2$ corresponding to the eigenvalues $\kappa_j(\kappa_j\pm
1)$. Eqs. (\ref{rad}) and (\ref{PP}) can be solved separating the
variables as,
\begin{equation}
f^{\pm}_{p,\kappa_j}(t_c,r)=\tau^{\pm}_p(t_c)
\rho_{p,\kappa_j}^{\pm}(r)\,,
\end{equation}
and finding  that the new functions must satisfy \cite{C4}
\begin{eqnarray}
\left(\pm i\partial_{t_c}+\frac{k}{t_c}
\right)\tau^{\pm}_p(t_c)&=&\pm
p \,\tau^{\mp}_p(t_c)\,,\\
\left(\pm
\partial_{r}+\frac{\kappa_j}{r} \right)\rho^{\pm}_{p,\kappa_j}(r)&=& p
\,\rho^{\mp}_{p,\kappa_j}(r)\,.
\end{eqnarray}
These equations have to be solved in terms of Bessel functions
\cite{AS} as in \cite{SHI}. The advantage of our method is to point
out that there is only one additional integration constant, $p$,
which is a continuous parameter with a precise physical meaning
(i.e., the scalar momentum).

However, our main problem is to find the normalized solutions with
respect to the scalar product (\ref{sp}). We specify that these
solutions are not square integrable since the spectrum of
$\vec{P}^2$ is continuous. Therefore, we must look for a system of
spinors $U_{p,\kappa_j, m_j}$ normalized in the generalized sense in
the scale of the scalar momentum $p$. First, we choose the radial
functions as in Ref. \cite{SHI},
\begin{equation}
\rho^{\pm}_{p,\kappa_j}(r)=\sqrt{pr}\,J_{|\kappa_j\pm
\frac{1}{2}|}(pr)\,.
\end{equation}
Furthermore, we observe that the functions $\tau^{\pm}$ must produce
a similar time modulation as in the case of the normalized plan wave
solutions presented above. Consequently, we denote
$\nu_{\pm}=\frac{1}{2}\pm ik$ and write the functions $\tau^{\pm}$
in terms of Hankel functions \cite{AS} as
\begin{equation}
\tau^{\pm}_p(t_c)= N\sqrt{-p t_{c}}\,e^{\pm\pi
k/2}H^{(1)}_{\nu_{\mp}}(-p t_{c})
\end{equation}
where $N$ is a normalization factor. Finally, summarizing these
results in the chart $\{t,r,\theta,\phi\}_*$ and matching the factor
$N$ we find the definitive form of the normalized spinors \cite{C4},
\begin{eqnarray}\label{sol}
&&U_{p,\kappa_j, m_j}(x)=\frac{p}{2} \sqrt{\frac{\pi}{\omega
r}}\,e^{-2\omega t}\left[e^{\pi
k/2}H^{(1)}_{\nu_-}(\textstyle{\frac{p}{\omega}} e^{-\omega t})
J_{|\kappa_j+\frac{1}{2}|}(pr)\Phi^+_{m_j,\kappa_j}(\theta,\phi)\right.\nonumber \\
&&~~~~~~~~~~~~~~~~~~~~~~~~~\left.+e^{-\pi
k/2}H^{(1)}_{\nu_+}(\textstyle{\frac{p}{\omega}} e^{-\omega t})
J_{|\kappa_j-\frac{1}{2}|}(pr)\Phi^-_{m_j,\kappa_j}(\theta,\phi)
\right]
\end{eqnarray}
that satisfy the Dirac equation and are common eigenspinors of the
operators $\vec{P}^2, \vec{J}^2, K$ and $J_3$. Taking into account
that \cite{LL}
\begin{equation}
\int_{0}^{\infty}\rho_{p,\kappa_j}(pr)
\rho_{p',\kappa_j}(p'r)dr=\delta(p-p')\,,
\end{equation}
and using the properties of Hankel functions mentioned in Appendix B
we obtain the orthonormalization rule
\begin{equation}
\left<U_{p,\kappa_j, m_j}, U_{p',\kappa'_{j'},
m'_j}\right>=\delta(p-p')\delta_{j,j'}\delta_{\kappa_j,\kappa'_{j}}\delta_{m_j,m'_j}\,.
\end{equation}
According to our conventions we can say that the particular
solutions (\ref{sol}) are of positive frequencies and, therefore,
these describe the quantum modes of the Dirac particles. The spinors
of negative frequencies, corresponding to antiparticles, will be
obtained using the charge conjugation,
\begin{equation}\label{conj1}
V_{p,\kappa_j, m_j}=(U_{p,\kappa_j, m_j})^{c}={\cal C}
(\overline{U}_{p,\kappa_j, m_j})^T \,.
\end{equation}
One can verify that these spinors have good orthonormalization
properties
\begin{equation}
\left<V_{p,\kappa_j, m_j}, V_{p',\kappa'_{j'},
m'_j}\right>=\delta(p-p')\delta_{j,j'}\delta_{\kappa_j,\kappa'_{j}}\delta_{m_j,m'_j}\,.
\end{equation}
and
\begin{equation}
\left<U_{p,\kappa_j, m_j}, V_{p',\kappa'_{j'},
m'_j}\right>=\left<V_{p,\kappa_j, m_j}, U_{p',\kappa'_{j'},
m'_j}\right>=0\,.
\end{equation}

\subsection{Quantization}

We obtained here two complete systems of orthonormalized spinors
which could be the starting point to the canonical quantization of
the Dirac field in moving frames. This procedure is simpler in the
case of the plane waves in the chart $\{t,\vec{x}\}_*$ where the
massive Dirac field reads
\begin{eqnarray}
\psi(t,\vec{x})&=&\psi^{(+)}(t,\vec{x})+\psi^{(-)}(t,\vec{x})\nonumber\\
&=&\int d^3 p \sum_{\lambda}\left[U_{\vec{p},\lambda}(x)
a(\vec{p},\lambda)+V_{\vec{p}, \lambda}(x)b^{\dagger}(\vec{p},
\lambda) \right]\,.\label{psiab}
\end{eqnarray}
The quantization can be done considering that the wave functions in
momentum representation, $a(\vec{p},\lambda)$ and
$b(\vec{p},\lambda)$, become field operators (so that $b^{*}\to
b^{\dagger}$).

We assume that the particle ($a$, $a^{\dagger}$) and antiparticle
($b$, $b^{\dagger}$) operators must fulfill the standard
anticommutation relations in the momentum representation, from
which the non-vanishing ones are
\begin{equation}\label{acom}
\{a(\vec{p},\lambda),
a^{\dagger}({\vec{p}}^{\,\prime},\lambda^{\prime})\}=
\{b(\vec{p},\lambda),
b^{\dagger}({\vec{p}}^{\,\prime},\lambda^{\prime})\}=
\delta_{\lambda\lambda^{\prime}}\delta^3
(\vec{p}-{\vec{p}}^{\,\prime})\,,
\end{equation}
since then the equal-time anticommutator takes the {\em canonical}
form
\begin{equation}
\{ \psi(t,\vec{x}),\, \overline{\psi}(t, \vec{x}^{\,\prime})\}=
e^{-3\omega t}\gamma^0 \delta^{3}(\vec{x}-\vec{x}^{\,\prime})\,,
\end{equation}
as it results from (\ref{compl}). In general, the partial
anticommutator functions,
\begin{equation}
\tilde S^{(\pm)}(t,t',\vec{x}-\vec{x}^{\,\prime})= i\{
\psi^{(\pm)}(t,\vec{x}),\, \overline{\psi}^{(\pm)}(t',
\vec{x}^{\,\prime})\}\,,
\end{equation}
and the total one $\tilde S=\tilde S^{(+)}+\tilde S^{(-)}$ are
rather complicated since for $t\not=t'$ we have no more identities
like (\ref{H3}) which should simplify their time-dependent parts.
In any event, these are solutions of the Dirac equation in both
their sets of coordinates and help one to write the Green
functions in usual manner. For example, from the standard
definition of the Feynman propagator \cite{SW1},
\begin{eqnarray}
&&\tilde S_F(t,t',\vec{x}-\vec{x}^{\,\prime})=
i\left<0\right|T[\psi(x)\overline{\psi}(x')]\left|0\right>\\
&&=\theta(t-t')\tilde S^{(+)}(t,t',\vec{x}-\vec{x}^{\,\prime})-
\theta(t'-t)\tilde S^{(-)}(t,t',\vec{x}-\vec{x}^{\,\prime})\,,
\end{eqnarray}
we find that
\begin{equation}
[E_{D}(x)-m]\tilde S_F(t,t',\vec{x}-\vec{x}^{\,\prime})=
-e^{-3\omega t} \delta^{4}(x-x')\,.
\end{equation}

Another argument for this quantization procedure is that the
one-particle operators given by the Noether theorem  have similar
structures and properties like those of the quantum theory of the
free fields in flat spacetime. Indeed, from Eq. (\ref{cq}) we
obtain the one-particle operators of the form (\ref{opo}) which
satisfy the Eqs. (\ref{algXX}). The diagonal one-particle
operators result directly using Eqs. (\ref{orto1})-(\ref{WUV}). In
this way we obtain the momentum components
\begin{equation}
{\bf P}^i=:\left<\psi,P^{i}\psi\right>:= \int d^3 p\,
p^i\sum_{\lambda}
\left[a^{\dagger}(\vec{p},\lambda)a(\vec{p},\lambda)
+b^{\dagger}(\vec{p},\lambda)b(\vec{p},\lambda)\right]
\end{equation}
and the helicity (or Pauli-Lubanski) operator
\begin{equation}
{\bf W}=:\left<\psi,W\psi\right>:= \int d^3 p  \sum_{\lambda}
p\lambda \left[a^{\dagger}(\vec{p},\lambda)a(\vec{p},\lambda)
+b^{\dagger}(\vec{p},\lambda)b(\vec{p},\lambda)\right]\,.
\end{equation}
The definition (\ref{opo}) holds for the generators of internal
symmetries too, including the particular case of  $X=1$ when the
bracket
\begin{equation}
{\bf Q}=\,:\left<\psi,\psi\right>:\,= \int d^3 p  \sum_\lambda
\left[a^{\dagger}(\vec{p},\lambda)a(\vec{p},\lambda)
-b^{\dagger}(\vec{p},\lambda)b(\vec{p},\lambda)\right]
\end{equation}
gives just the charge operator corresponding to the internal
$U(1)$ symmetry of the action (\ref{action}) \cite{TH,BD}. It is
obvious that all these operators are self-adjoint and represent
the generators of the external or internal symmetry
transformations of the quantum fields \cite{SW1}. The conclusion
is that, for fixed mass and spin, the helicity state vectors of
the Fock space defined as common eigenvectors of the set $\{{\bf
Q}, {\bf P}^i, {\bf W}\}$ form a complete system of
orthonormalized vectors in generalized sense, i.e. the helicity
basis.

The Hamiltonian  operator ${\bf H}=:\left<\psi,H\psi\right>:$ is
conserved but is not diagonal in this basis since it does not
commute  with ${\bf P}^i$ and ${\bf W}$ as it follows from the
commutation relations (\ref{cHP}) and the properties
(\ref{algXX}). Its form in momentum representation can be
calculated using the identity
\begin{equation}
H\,U_{\vec{p},\lambda}(t,\vec{x})=-i\omega
\left(p^i\partial_{p^{i}}+\frac{3}{2}\right)
U_{\vec{p},\lambda}(t,\vec{x})\,,
\end{equation}
and the similar one for $V_{\vec{p},\lambda}$, leading to
\begin{equation}
{\bf H}=\frac{i\omega}{2}\int d^3 p\,p^i\sum_{\lambda}\left[
a^{\dagger}(\vec{p},\lambda)\stackrel{\leftrightarrow}{\partial}_{p^{i}}
a(\vec{p},\lambda) +
b^{\dagger}(\vec{p},\lambda)\stackrel{\leftrightarrow}{\partial}_{p^{i}}
b(\vec{p},\lambda) \right]
\end{equation}
where the derivatives act as
$f\stackrel{\leftrightarrow}{\partial}h=f\partial h -(\partial f)
h$. Hereby it results the expected  behavior of ${\bf H}$ under
the space translations of $\tilde E(3)$ which transform the
operators $a$ and $b$ according to (\ref{tri}). Moreover, it is
worth pointing out that the change of the phase factors
(\ref{gaugeUV}) associated with the transformations
\begin{equation}\label{gaugeab}
a(\vec{p},\lambda)\to
e^{-i\chi(\vec{p})}a(\vec{p},\lambda)\,,\quad
b(\vec{p},\lambda)\to e^{-i\chi(\vec{p})}b(\vec{p},\lambda)
\end{equation}
leave invariant the operators $\psi$, ${\bf Q},\,{\bf P}^{i}$ and
${\bf W}$ as well as the equations (\ref{acom}), but transform the
Hamiltonian operator,
\begin{equation}\label{trah}
{\bf H}\to {\bf H} + \omega\int d^3 p\, [p^i
\partial_{p^i}\chi(\vec{p})] \sum_\lambda
\left[a^{\dagger}(\vec{p},\lambda)a(\vec{p},\lambda)
+b^{\dagger}(\vec{p},\lambda)b(\vec{p},\lambda)\right]\,.
\end{equation}
This remarkable property may be interpreted as a new type of gauge
transformation depending on  momentum instead of coordinates. Our
preliminary calculations indicate that this gauge may be helpful
for analyzing the behavior of the theory near $\omega\sim 0$.

In the simpler case of the left-handed massless field with the
fundamental spinor solutions (\ref{n1}) and (\ref{n2}) we obtain
similar results and we recover the standard rule of the neutrino
polarization.

\subsection{Further questions}

The quantum theory presented above opens many new interesting
problems. One of them is if in the dS geometry there exists an
orbital analysis analogous to the Wigner theory of induced
representations of the Poincar\' e group \cite{TH}. This is
necessary if we want to understand the meaning of the rest frames
(of the massive particles) in the dS spacetime and to find the
``booster" mechanisms changing the value of $p$ or even giving rise
to waves of arbitrary momentum from those with $\vec{p}=0$. We
believe that this theory may be done starting with the orbital
analysis in $M^5$ since this helps us to find $SO(4,1)$-covariant
definitions for our basic operators on $M$ \footnote{For plane wave
spinors of a Dirac theory in $M^5$ see Ref. \cite{TTT}}. More
precisely, for each momentum $q\in M^5_{q}$ we can write a
five-dimensional momentum operator $P(q)$ of components
$P^A(q)=\eta^{AC}q^B X_{(BC)}$ while a generalized five-dimensional
Pauli-Lubanski operator in $M$ has to be defined by
$W_{A}=-\frac{1}{8}\varepsilon_{ABCDE}X^{(BC)}X^{(DE)}$ . Then it is
clear that for the representative momentum $\hat
q=(\omega,0,0,0,-\omega)$ of the orbit $q^2=0$, associated to the
little group $E(3)\subset SO(4,1)$, we recover our operators $P(\hat
q)=(H, \vec{P}, -H)$ and $W=\hat q^{A}W_A$ as given by (\ref{Gi}),
(\ref{Gip}) and (\ref{PL}), respectively. In this way one may
construct  generalized Wigner  representations of the group $S(M)$
in spaces of spinors depending on momentum.

In other respects, it is important to investigate the physical
consequences of the transformation laws of the main observables of
this theory and to point out the role and significance of the
transformations (\ref{gaugeab}) and (\ref{trah}). Of a particular
interest could be the study of the influence of the dS gravitational
field on the energy measurements since these are affected by the
uncertainty relations $\Delta H \Delta P^i \ge \omega
|\left<P^i\right>|/2$ due to the commutations relations (\ref{cHP}).
Of course, for very small values of $\omega$ it is less probable
that these produce observable effects in local measurements, the
spacetime appearing then as a flat one.

Other problems which could appear in further investigations of the
Dirac free field seem to be rather technical, e.g. the properties of
commutator and Green functions, calculation of the action of more
complicated conserved operators, evaluation of the inertial effects
etc.. However, in our opinion, the next important step from the
physical point of view would be to construct a similar theory for
the free electromagnetic field, completing thus the framework one
needs for developing the perturbative QED in the dS spacetime.

\section{Concluding remarks}

The methods we used here for deriving solutions of the Dirac
equation on dS and AdS spacetimes can be generalized to any
dimensions. We have shown that our Cartesian tetrad gauge in
Cartesian coordinates can be generalized to any central manifold
$M_{1+d}$ having $d$ space dimensions \cite{C5,C6}. In this gauge
the Dirac equation takes a suitable form which allows one to
separate the angular variables of the generalized spherical
coordinates according to the general method proposed in Ref.
\cite{XYGu}. After this separation one is left with two radial
equations that can be analytically solved in some interesting cases.

In this context we have found that the massive Dirac particles
moving freely on ${\rm CAdS}_{1+d}$ backgrounds have the discrete
and equidistant energy spectra \cite{C5}
\begin{equation}\label{(enlev1)}
E_{n}=m+\omega\left(n+\frac{d}{2}\right)\,,\quad n=0,1,2,....\,,
\end{equation}
generalizing thus Eq. (\ref{(enlev)}) to any $d$. These spectra have
the ground state energies $E_0=m+\omega\frac{d}{2}$ which are in
accordance with the scaling dimensions of the spinor operators of
CFT which in our notations read $\Delta_{sp}=k+\frac{d}{2}$ with
$k=m/\omega$ \cite{CFT1}.

We note  that the scalar field on ${\rm CAdS}_{1+d}$  has also
equidistant energy levels, \cite{BL,C7}
\begin{equation}
E_n^{sc}
=\sqrt{m^2+\frac{d^2}{4}\omega^2}+\omega\left(n+\frac{d}{2}\right)
\,,\quad n=0,1,2,... ~ \,,
\end{equation}
with the same energy quanta as in Eq. (\ref{(enlev1)}) but having
another ground state energy. This difference is due to the fact that
in ${\rm AdS}_{1+d}$ spacetimes the free Dirac equation is no more
the square root of the Klein-Gordon equation with the same mass. The
ground state energy defines the conformal dimension
\begin{equation}
\Delta_{sc}=\sqrt{k^2+\frac{d^2}{4}}+\frac{d}{2}
\end{equation}
of the scalar field  in CFT \cite{EW}.

Another interesting case is of the ${\Bbb R}\times S^d$ manifolds
defined as $d$-dimensional spheres of radius $1/\omega$ with the
time trivially added. The metrics of these manifolds differ from the
${\rm AdS}_{1+d}$ ones through conformal transformations. The
massive Dirac perturbations on ${\Bbb R}\times S^d$ manifolds have
also discrete energy spectra \cite{C6},
\begin{equation}\label{(enlev2)}
E_{n}=\left[
m^2+\omega^2\left(n+\frac{d}{2}\right)^2\right]^{\frac{1}{2}}\,,\quad
n=0,1,2,....\,,
\end{equation}
which are no longer equidistant if $m\not= 0$. However,  if the
Dirac particle is massless ($m=0$) then the spectra (\ref{(enlev1)})
and (\ref{(enlev2)}) coincide . This is because the massless Dirac
equation is invariant under conformal transformations \cite{BD}. In
fact, the massless Dirac perturbations have the same energy
spectrum,
\begin{equation}\label{(enlev0)}
E_{n}^0=\omega\left(n+\frac{d}{2}\right)\,,\quad n=0,1,2,....\,,
\end{equation}
on all of the manifolds whose metrics are conformal transformations
of the ${\rm AdS}_{1+d}$ or ${\Bbb R}\times S^d$ ones \cite{CaH}.

Finally, we would like to hope that our approach based on external
symmetries could be an argument for a general tetrad gauge covariant
theory of the quantum fields with spin in which the quantum modes
may be defined by complete sets of commuting operators in such a
manner that the procedure of second quantization should be
independent on the frames one uses.

\appendix

\subsection*{Appendix A: SO(4,1) transformations and dS isometries}

We have seen that the dS manifold, $M$, is a hyperboloid embedded in
$M^5$. The metric $\eta=(1,-1,-1,-1,-1)$ of $M^5$ is invariant under
the coordinate transformations $Z^A\to {^5\!\Lambda}^{A\,\cdot}
_{\cdot\,B}Z^B$ where $^5\!\Lambda\in SO(4,1)$. Each coordinate
transformation give rise to an isometry of the group $I(M)$ which
can be calculated in the local chart $\{t_{c},\vec{x}\}_*$ using
Eqs. (\ref{Zx}).  We remind the reader that the basis generators
$^5\! X_{AB}$ of the fundamental (linear) representation of
$SO(4,1)$, carried by $M^5$, have the matrix elements
\begin{equation}
(^5\! X_{AB})^{C\,\cdot}_{\cdot\,D}=i\left(\delta^C_A\, \eta_{BD}
-\delta^C_B\, \eta_{AD}\right)\,.
\end{equation}

The transformations of $SO(3)\subset SO(4,1)$ are simple rotations
of $Z^i$ and $x^i$ which transform alike since this symmetry is
global. For the other transformations generated by $H,\, P^i$ and
$N^i$ the linear transformations in $M^5$ and the isometries are
different. Those due to $H$,
\begin{equation}
e^{-i\xi_{H}{^5\!H}}\,:\quad
\begin{array}{lcl}
Z^0&\to&Z^0 \cosh\alpha-Z^5 \sinh\alpha \\
Z^5&\to&-Z^5 \sinh\alpha+Z^0 \cosh\alpha \\
Z^i&\to&Z^i
\end{array}
\end{equation}
where $\alpha=\omega\xi_{H}$, produce the dilatations $t_{c}\to
t_{c}e^{\alpha}$ and $x^i\to x^i e^{\alpha}$, while the
transformations
\begin{equation}
e^{-i\vec{\xi}_{P}\cdot {^5\!\vec{P}}}\,:\quad
\begin{array}{lcl}
Z^0&\to&Z^0 +\omega\,\vec{\xi}_{P}\cdot\vec{Z}+\frac{1}{2}\,\omega^2
{{\xi}_P}^2\,(Z^0+Z^5) \\
Z^5&\to&Z^5 -\omega\,\vec{\xi}_{P}\cdot\vec{Z}-\frac{1}{2}\,\omega^2
{{\xi}_P}^2\,(Z^0+Z^5) \\
Z^i&\to&Z^i+\omega\,\xi_P^i\,(Z^0+Z^5)
\end{array}
\end{equation}
give the space translations $ x^i\to x^i +\xi_P^i$ at fixed $t_c$.
More interesting are the transformations
\begin{equation}
e^{-i\vec{\xi}_{N}\cdot {^5\!\vec{N}}}\,:\quad
\begin{array}{lcl}
Z^0&\to&Z^0 -\vec{\xi}_{N}\cdot\vec{Z}+\frac{1}{2}\,
{{\xi}_N}^2\,(Z^0-Z^5) \\
Z^5&\to&Z^5 -\vec{\xi}_{N}\cdot\vec{Z}+\frac{1}{2}\,
{{\xi}_N}^2\,(Z^0-Z^5) \\
Z^i&\to&Z^i-\xi_N^i\,(Z^0-Z^5)
\end{array}
\end{equation}
which lead to the isometries
\begin{eqnarray}
t_c&\to&\frac{t_c}{1-2\omega\, \vec{\xi}_{N}\cdot\vec{x}
-\omega^2{{\xi}_N}^2\,({t_c}^2-r^2)} \\
x^i&\to&\frac{x^i+\omega\xi_N^i\, ({t_c}^2-r^2)} {1-2\omega\,
\vec{\xi}_{N}\cdot\vec{x} -\omega^2{{\xi}_{N}}^2\,({t_c}^2-r^2)}\,.
\end{eqnarray}
We denoted here ${\xi_P}^2=(\vec{\xi}_P)^2$ and
${\xi_N}^2=(\vec{\xi}_N)^2$.

\subsection*{Appendix B: Some properties of Hankel functions}

According to the general properties of the Hankel functions
\cite{AS}, we deduce that those used here,
$H^{(1,2)}_{\nu_{\pm}}(z)$, with $\nu_{\pm}=\frac{1}{2}\pm i k$
and $z\in \Bbb R$, are related among themselves through
\begin{equation}\label{H1}
[H^{(1,2)}_{\nu_{\pm}}(z)]^{*} =H^{(2,1)}_{\nu_{\mp}}(z)\,,
\end{equation}
satisfy the equations
\begin{equation}\label{H2}
\left(\frac{d}{dz}+\frac{\nu_{\pm}}{z}\right)H^{(1)}_{\nu_{\pm}}(z)
=  i e^{\pm \pi k} H^{(1)}_{\nu_{\mp}}(z)
\end{equation}
and the identities
\begin{equation}\label{H3}
e^{\pm \pi k} H^{(1)}_{\nu_{\mp}}(z)H^{(2)}_{\nu_{\pm}}(z) +
e^{\mp \pi k}
H^{(1)}_{\nu_{\pm}}(z)H^{(2)}_{\nu_{\mp}}(z)=\frac{4}{\pi z}\,.
\end{equation}

\end{document}